\documentclass[a4paper,10pt]{article}
\usepackage{fullpage}
\usepackage{color}
\usepackage{graphicx}
\usepackage[english]{babel}
\usepackage{float}
\usepackage{amsmath,amssymb,dsfont,authblk,amsfonts}
\usepackage{braket} 
\usepackage{cite}
\usepackage{hyperref}
\usepackage{subcaption} % subfigures, subcaptions: mehrere Bilder nebeneinander

\usepackage{titlesec}
\titlelabel{\thetitle.\quad}

\newcommand{\norm}[1]{\left\lVert#1\right\rVert}
\newcommand{\abs}[1]{\ensuremath{\left\vert#1\right\vert}}
\newcommand{\HA}{{\text{HA}}}
\newcommand{\HV}{{\text{HV}}}

\newcommand{\unit}{1\!\!1}

\renewcommand{\Re}{{\text{Re }}}
\renewcommand{\Im}{{\text{Im }}}
\newcommand{\Tr}{{\text{Tr}}}
\newcommand{\Exp}{{\text{Exp}}}
\newcommand{\Det}{{\text{Det }}}
\newcommand{\Per}{{\text{Per }}}
\newcommand{\Arccos}{{\text{arccos }}}
\newcommand{\Vol}{{\text{Vol }}}
\newcommand{\sgn}{{\text{sgn }}}

\newcommand{\vc}[1]{{\boldsymbol #1}}

\numberwithin{equation}{section}

\begin{document}
\title{Hilbert space average of transition probabilities}
\author{Nico Hahn, Thomas Guhr, Daniel Waltner}
\affil{Fakult\"at f\"ur Physik, Universit\"at Duisburg-Essen, Lotharstra\ss e 1, 47048 Duisburg, Germany.}
\maketitle

\begin{abstract}
The typicality approach and the Hilbert space averaging method as its technical manifestation are important concepts of quantum statistical mechanics. Extensively used for expectation values we extend them in this paper to transition probabilities. In this context we also find that the transition probability of two random uniformly distributed states is connected to the spectral statistics of the considered operator. Furthermore, within our approach we are capable to consider distributions of matrix elements between states, that are not orthogonal. We will demonstrate our quite general result numerically for a kicked spin chain in the integrable resp. chaotic regime.
\end{abstract}

\section{Introduction}

Consider a set of pure states $\{ \ket{\phi} \}$ initially featuring very similar expectation values concerning an observable $M$. Then dynamical quantum typicality refers to the situation in which the majority of these states will also feature very similar expectation values to the same or another observable at any later time \cite{Bartsch}. This statement holds true under quite general conditions as it is not even dependent on the details of the dynamics. A necessary requirement is that the dimension $N$ of the finite dimensional Hilbert space $\mathcal{H}$ is large. In \cite{Reimann1} the construction of these sets was generalized and sufficient conditions were explored. It was found that besides a large Hilbert space dimension the preset expectation value $\braket{\phi |M| \phi}$ as well as the spectrum of $M$ play an important role.

The tool to develop these findings is the Hilbert space averaging method \cite{Gemmer}. With its aid one can establish the average of a function depending on pure states, e.g. the expectation value of some arbitrary observable. The standard deviation as the square root of the Hilbert space variance describes the width of the distribution of the states around this value and is found to be small if certain conditions are met \cite{Reimann1}. Using these methods it was possible to get more insight into the eigenstate thermalization hypothesis (ETH) \cite{Reimann2,Steinigeweg3, Khodja, Steinigeweg4}, that strives to explain the thermalization of isolated quantum systems. Furthermore the resulting identities have been utilized as a computational scheme, e.g. to calculate spin-correlations \cite{Steinigeweg1,Steinigeweg2,Elsayed,Richter} as well as in the case of out-of-time-ordered correlations (OTOC's) \cite{Prosen2}, which are able to diagnose many- body localized phases \cite{Chen}. The intention of the present work is to extend the Hilbert space averaging method and open it up to new possibilities by studying the transition probabilities $\abs{\braket{\chi|A|\psi}}^2$ of an arbitrary operator $A$.

On the technical level this averaging procedure is closely connected to the concept of random matrix theory \cite{Guhr, Mehta}.
This method substitutes the matrices, relevant for the description of the system (e.g. the Hamiltonian, scattering matrix etc.), by  random matrices. 
It was applied quite successfully to systems sharing a sufficiently high amount of complexity. 
The only information about the system entering the model are its symmetries, revealing universal system features exclusively depending on the symmetry class. In previous works, where concepts of random matrix theory have been applied to study transition probabilities \cite{Barbosa, Alhassid1, Ullah1, Ullah2} the components of the states were drawn from a Gaussian distribution with zero mean, implying statistical orthogonality of the states in high dimensions. Thus, these states are able to model the eigenstates of an observable and the transition probability between those eigenstates. A well known result of this approach is the Porter-Thomas distribution \cite{PorterThomas, Alhassid2}, which determines the statistics of the transition probabilities in the case of time reversal symmetry. However, in this paper we follow a new route, where we explicitly choose the overlap of the deployed states, putting the assumption of statistical orthogonality aside. The situation of nonorthogonal states is relevant, when states, that are not eigenstates of the Hamiltonian or another observable, important for the system description, are prepared. For example within the doorway mechanism (see \cite{Kohler1, Kohler2} and references therein) a distinct state (the so-called doorway state), that is not an eigenstate of the Hamiltonian, but is distinguished by another feature (e.g. collective motion in many-body systems) is coupled to a background of states, that are not orthogonal to the doorway state. This mechanism has far reaching applications in nuclear physics (giant multipole resonance) \cite{Zelevinsky}, molecular physics \cite{Kawata, Lombardi} and mesoscopic physics \cite{Aberg, Hussein}.

The outline of the paper is as follows: in section \ref{section2} we will introduce the Hilbert space averaging method and give a short review on the concept of quantum typicality, focusing especially on the construction of typical states and on the aforementioned conditions of dynamical quantum typicality. In section \ref{section3} we will shortly introduce the model on which we present our findings. In section \ref{section4} we approach the Hilbert space average of transition probabilities, first without imposing any further condition onto the states, and then after fixing the initial overlap of the involved states. We demonstrate and interpret our results with the aid of numerical calculations in section \ref{section5}. More involved analytical calculations are relegated to the \hyperref[Appendix]{appendix}.

\section{Framework} \label{section2}

We restrict ourselves to a finite dimensional Hilbert space $\mathcal{H} = \mathbb{C}^N$ with dimension $N$. In practice this will be due to the nature of the system or because of conserved quantities, e.g. the Hilbert space could be the energy shell of an isolated system. We now give an overview of the concept of quantum typicality.

We want to find the average value of a function $f: \mathcal{H} \rightarrow \mathbb{C}$. This is the Hilbert space average \cite{Gemmer}
\begin{equation} \label{2HADef}
\HA\left[f(\psi)\right] = \frac{\int d[\psi] f(\psi) \delta (\braket{\psi|\psi} - 1)}{\int d[\psi] \delta (\braket{\psi|\psi} - 1)}.
\end{equation}
Thus, we choose normalized vectors as representative for the state, which allows us to restrict our integration to the unit sphere in $\mathcal{H} = \mathbb{C}^N$. Then we normalize this integral by the volume of the unit sphere. 

We note that for a normalized vector there is still a relative phase $e^{i \lambda}$ left as a degree of freedom. However, the functions, treated on the next pages, fulfill $\forall \lambda \in [0,2\pi): f(\psi) = f( e^{i \lambda} \psi)$ (in other words they are $U(1)$ invariant) and our normalization makes up for that multiple occurrence of identical states in the integral over the sphere $S^{2N-1}$. The latter can be seen as a fibre bundle $S^{2N-1} \sim \mathbb{C}P^{N - 1} \times U(1)$, whose base manifold is the complex projective space $\mathbb{C}P^{N - 1}$, that is the space of all rays in $\mathbb{C}^N$ \cite{Bengtsson}. Likewise the Hilbert space variance can be defined as
\begin{equation}
\HV [f(\psi)] = \HA\left[f^2(\psi) \right] - \HA^2\left[f(\psi)\right].
\end{equation}

A first important application is the expectation value of an operator $M$.  As shown in \cite{Gemmer} this can be obtained without using the explicit form of \eqref{2HADef}. Because the integration regime is invariant under the action of the unitary group $U(N)$ (this is also known as Haar measure \cite{Bengtsson}) we can choose an arbitrary orthonormal basis $\{\ket{i}\}$ to represent our states and then use the linearity of the Hilbert space average
\begin{equation}
\HA\left[\braket{\psi|M| \psi}\right] = \sum_{i,j} M_{ij}\ \HA\left[\psi_i^* \psi_j\right]
\end{equation}
with the coordinates $\psi_i = \braket{i|\psi}$. The unit sphere is invariant under $\psi_i \rightarrow - \psi_i$ for all $i$ and therefore the average will vanish for $i \neq j$. Now we can use the normalization of our states
\begin{equation}
\HA\left[\sum_i \abs{\psi_i}^2\right] = \sum_i \HA\left[\abs{\psi_i}^2\right] = 1
\end{equation}
to conclude
\begin{equation}
\HA\left[\abs{\psi_i}^2\right] = \frac{1}{N}
\end{equation}
and 
\begin{equation}
\HA\left[\braket{\psi|M|\psi}\right] = \frac{\Tr\ M}{N}. \label{2HAExp}
\end{equation}
The Hilbert space variance of the expectation value is \cite{Gemmer}
\begin{equation} \label{2HV}
\HV\left[\braket{\psi|M|\psi}\right] = \frac{1}{N+1} \left(\frac{\Tr\ M^2}{N} - \frac{\Tr^2 M}{N^2}\right).
\end{equation}
A sharp Hilbert space average, in the sense that the standard deviation drops faster in $N$ than the average, can be reached by demanding constant spectral moments $\Tr\ M/N$ and $\Tr\ M^2/N$ \cite{Bartsch}. The calculation of higher moments of $\braket{\psi|M|\psi}$ is more complicated. In appendix \ref{Aaveragematrixelements} we present a formula, that achieves this for positive integer powers.

Next we will review the typicality results of \cite{Reimann1} and choose $M$ as an observable. Reimann constructs a statistical operator 
\begin{equation} \label{2RhoReimann}
\rho (m,M) = \frac{1}{N} \frac{1}{1+ y (m - M)},
\end{equation}
where $m$ is the desired expectation value of $\rho (m,M)$ concerning $M$. The parameter $y \in \mathbb{R}$ is chosen so that the normalization $\Tr\ \rho = 1$ and $\Tr\ M \rho = m$ are fulfilled, and is determined by the roots of a rational function. The states
\begin{equation}
\ket{\phi}:=\Lambda \ket{\psi}, \label{2PhiDef}
\end{equation}
where we use the abbreviation $\Lambda = \sqrt{N \rho}$, exhibit the desired preset expectation value $m$ as their Hilbert space average
\begin{equation}
\HA\left[\braket{\phi|M|\phi}\right]= \HA\left[\braket{\psi|\Lambda M \Lambda |\psi}\right] = \Tr\ M \rho = m.   \label{2HAA}
\end{equation}
They are not necessarily normalized, but in average they are
\begin{equation} \label{2HAPhiNorm}
\HA\left[\braket{\phi|\phi}\right]= \HA\left[\braket{\psi|\Lambda^2|\psi}\right] = \Tr\ \rho = 1.
\end{equation}
The Hilbert space variance indicates how much one can trust these results in the case of an individual $\ket{\phi}$. The purity
\begin{equation}
P = \Tr\ \rho^2
\end{equation}
is an upper bound for the variances
\begin{equation} \label{2HVPhiNorm}
\HV\left[\braket{\phi|\phi}\right] = \frac{NP - 1}{N+1} \leq P \qquad \text{and} \qquad \HV\left[\braket{\phi|M|\phi}\right] \leq \norm{M}^2 P,
\end{equation}
where the operator norm $\norm{M}$ is defined as the absolute value of the largest eigenvalue in modulus. Therefore one achieves typicality if the purity is low $P \ll 1$. While a large Hilbert space is a prerequisite for low purity, due to \eqref{2RhoReimann} $P$ is also strongly dependent on the observable $M$ and the expectation value $m$. As an example consider the magnetization in $z$-direction
\begin{equation} \label{2Magnetization}
M_z = \sum_{i=1}^n \sigma_i^z
\end{equation}
of an $n$-particle spin-1/2 chain (cf. section \ref{section3}). In this case the eigenvalues of the local spin observables $\sigma_i^z$ are $\pm 1$. Therefore the eigenvalue density of $M_z$ is a binomial distribution, i.e. there are only few eigenstates near the maximal resp. minimal magnetization of $\pm n$, while the majority is located around a magnetization of zero in the center. In fact, there is only a single state that has the maximal resp. minimal magnetization as its expectation value, making it highly untypical. This leads to a high purity of statistical operators with expectation values near those values. It is exactly the other way around for statistical operators with $\Tr\ \rho M_z \approx 0$.

Finally we note that the above equations remain valid if $M$ is replaced by any other observable $B(t)$, that does not fulfill \eqref{2HAA}. Thus, dynamical quantum typicality is retrieved, namely that states $\{\ket{\phi}\}$ initially featuring $\braket{\phi|M|\phi} \approx m$ with small variance will also center around a common expectation value for another observable $B$ with small variance. As the deployed observables enter only in terms of their operator norm, the upper bound of the variance \eqref{2HVPhiNorm} is time independent for unitary time evolution.

\begin{figure}[t]
\centering
\includegraphics[width=.9\linewidth]{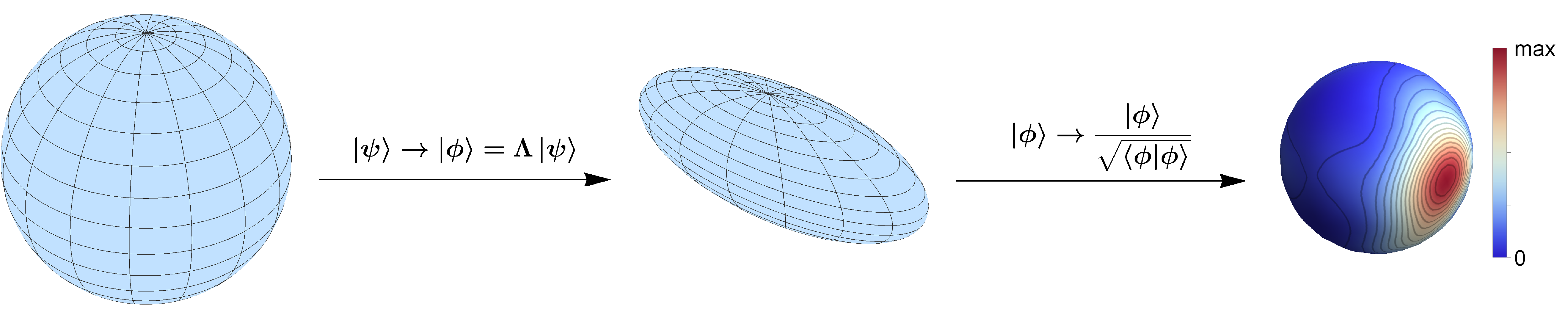}
\caption{Exemplification of the protocol, described in the main text. The density plot on the right is generated by interpolating the data of 100 uniformly sampled vectors after applying $\Lambda$ and renormalization. The colour code represents the likelihood to sample one of these vectors in the corresponding area.}
\label{ProjSphere}
\end{figure}

We want to give a geometrical interpretation of what is done here, illustrated in Fig.\ \ref{ProjSphere} for $N=3$. We uniformly sampled 100 vectors (in $\mathbb{R}^3$) on the unit sphere and applied $\Lambda$ with random positive eigenvalues of $\rho$, so that $\Tr\ \rho = 1$. Because $\Lambda$ is positive and Hermitian one can comprehend it as the deformation of the former integration regime (the unit sphere) into an ellipsoid. The new states $\{ \ket{\phi} \}$ are still uniformly distributed, but now over the ellipsoid. This ellipsoid is then described by the quadric $\braket{\psi|\Lambda^{-2}|\psi} - 1 = 0$. A subsequent normalization of each state will generate a nonuniform probability distribution on the unit sphere, which is given by the norm of the states on the surface of the ellipsoid $p (\psi) := \norm{\Lambda \ket{\psi}}$. In practice we interpolated our data \cite{Diggle} to generate the density plot shown in the figure. We want to make a short remark about \eqref{2PhiDef}. Inserting the fully mixed state $\rho_\text{{m}} = \unit/N$ (the one with the lowest purity) yields the uniform distribution from which we drew the states $\ket{\psi}$.

\section{Kicked Ising Chain}\label{section3}

Now we introduce the model, on which we want to present the findings of the next section. The kicked Ising chain (KIC) is a many-body quantum system that entails rich dynamics, reaching from integrable to fully chaotic. It consists of a closed ring of $n$ spins (here spin-1/2 thus the Hilbert space dimension is $N = 2^n$) with nearest neighbour interaction and an on site magnetic field. Additionally the system is periodically kicked by a transverse magnetic field leading to a discretized time evolution. The Hamiltonian is thus a sum of two parts
\begin{equation}
H = H_I + H_K \sum_{\tau=-\infty}^\infty \delta(t-\tau).
\end{equation}
The interaction part is
\begin{equation}
H_I = \sum_{i=1}^n \left( J \sigma_i^z \sigma_{i+1}^z + h\ \sigma_i^z \right),
\end{equation}
where $J$ is the coupling strength and $h$ the on site magnetic field. The kick part is
\begin{equation}
H_K = \sum_{i=1}^n b\ \sigma_i^x
\end{equation}
with the transverse magnetic field $b$. In the present case of a spin-1/2 chain, $\sigma_i^{\alpha}$ with $\alpha \in \{x,y,z\}$ are the Pauli matrices acting at site $i$. Due to the boundary condition we have $\sigma^\alpha_i = \sigma^\alpha_{i+n}$. The time evolution operator for one period (Floquet operator) also splits in two parts and is given by
\begin{equation} \label{3Floquet}
U= \mathcal{T}\ \Exp \left[ -i \int_0^1 H(t) dt \right] = U_I U_K
\end{equation}
with
\begin{equation} \label{3IsingKickpart}
U_I=\Exp \left[-i  H_I \right],\qquad U_K=\Exp \left[-i H_K \right]  
\end{equation}
and the time ordering operator $\mathcal{T}$. Here we use units in which $\hbar = 1$. The system becomes integrable for $b=0$ as the Hamiltonian will be diagonal in the basis of $\sigma^z$ and also if $h=0$. In the latter case it can be mapped onto a chain of noninteracting spinless fermions via a Jordan-Wigner transformation \cite{Lieb, Akila}. 

In quantum chaos \cite{Stoeckmann,Haake} the spectral statistics of the operators governing the dynamics, i.e. the Hamiltonian and the time evolution operator, is used to identify regular and chaotic behaviour. In this context the spectral form factor
\begin{equation}
K(T) = \frac{\abs{\Tr\ U^T}^2}{N}
\end{equation}
is a popular quantity. In recent years the KIC form factor was well examined \cite{Akila, Prosen1, Braun}. We will reconsider it in section \ref{section5}, where we find a connection between spectral statistics and the Hilbert space average.

\section{Hilbert space average of transition probabilities} \label{section4}

We now turn to transition probabilities $\abs{\braket{\chi|A|\psi}}^2$ describing the probability to find $\ket{\psi}$ in $\ket{\chi}$ after applying an arbitrary operator $A$. Assuming again that $\ket{\psi}$ is random uniformly distributed the Hilbert space average evaluates to
\begin{equation}
\HA_{\psi}\left[\abs{\braket{\chi | A | \psi}}^2 \right] = \HA_{\psi}\left[\braket{\psi| \left( A^\dag \ket{\chi}\bra{\chi}A \right) |\psi} \right] = \frac{\Tr\ A^\dag \ket{\chi}\bra{\chi}A}{N} = \frac{\braket{\chi| A A^\dag |\chi}}{N}.  \label{4HAIndependent}
\end{equation}
Here we used \eqref{2HAExp} and henceforth we indicate which state will be averaged by a subscript if ambiguities cannot be excluded. The order of $A$ and its adjoint operator on the right hand side changes if $\ket{\psi}$ and $\ket{\chi}$ on the left hand side are interchanged. This can be disregarded if $A$ commutes with its adjoint which is the case for Hermitian or unitary operators.

To understand this result, it helps to consider, for the time being, a unitary operator $A$, which does not change the relative orientation of the states and thus has no effect on their distribution. Then the average probability to find $\ket{\psi}$ parallel to one of the $N$ axes of the vector space is 
\begin{equation}
\HA_{\psi}\left[\abs{\braket{\chi | A | \psi}}^2\right] = \frac{1}{N}, \label{4AverageOverlap}
\end{equation}
leading to their statistical orthogonality in the case of large $N$. Returning to a general operator $A$ we can average the remaining state $\ket{\chi}$ and arrive at
\begin{equation}
\HA_{\psi, \chi}\left[\abs{\braket{\chi | A | \psi}}^2\right] = \HA_{\chi}\left[ \frac{\braket{\chi| A A^\dag |\chi}}{N}  \right] = \frac{\Tr\ AA^\dag}{N^2}.
\end{equation}
Endowing $\ket{\psi}$ as well as $\ket{\chi}$ with $\Lambda = \sqrt{N\rho}$ and $\Lambda' = \sqrt{N\rho'}$, where $\rho$ and $\rho'$ are arbitrary statistical operators, like in \eqref{2PhiDef} we find
\begin{equation}
\HA_{\psi,\chi}\left[\abs{\braket{\chi |\Lambda' A \Lambda| \psi}}^2 \right] =  \Tr\ \rho' A \rho A^\dag
\end{equation}
and especially for $\rho' = \rho$ and $A=\unit$
\begin{equation}\label{4PasEffDim}
\HA_{\psi,\chi}\left[\abs{\braket{\chi | \Lambda^2| \psi}}^2 \right] = P.
\end{equation}
Comparing \eqref{4AverageOverlap} and \eqref{4PasEffDim} this motivates the effective dimension
\begin{equation}
d_{\text{eff}} := \frac{1}{P}
\end{equation}
used by various authors \cite{Reimann2,Popescu}, which describes the variety of different states appearing in the mixture $\rho$. Applying \eqref{2HV} for the corresponding Hilbert space variances yields
\begin{align}
\HV_{\psi}\left[\abs{\braket{\chi | A | \psi}}^2\right] & = \frac{N-1}{N^2 (N+1)} \braket{\chi|AA^\dag|\chi}^2
\\
\HV_{\psi,\chi}\left[\abs{\braket{\chi | A | \psi}}^2\right] & = \frac{N-1}{N^3 (N+1)^2} \left( \Tr \left( AA^\dag \right)^2 + \Tr^2 AA^\dag \right). \nonumber
\end{align}

In the remaining part of the paper we will put the statistical orthogonality, revealed in \eqref{4AverageOverlap}, aside by fixing the overlap of $\ket{\psi}$ and $\ket{\chi}$ to an arbitrary complex number $z = \braket{\chi|\psi}$ within the unit circle ($\abs{z} \leq 1$). We do this by adding the corresponding $\delta$-function into the Hilbert space average. As its argument is complex, it has to be understood in terms of a product of two $\delta$-functions concerning the real and imaginary parts. Initially, we want to keep $\ket{\chi}$ fixed, which leads to the following expression
\begin{align} \label{4HAFixedResult}
\HA_{\psi}\left[\abs{\braket{\chi |A| \psi}}^2 \delta (\braket{\chi | \psi} - z)\right] &=  \frac{1 - \abs{z}^2}{N - 1}\braket{\chi|AA^\dag|\chi} + \frac{N\abs{z}^2 - 1}{N - 1} \abs{\braket{\chi|A|\chi}}^2.
\end{align}
A detailed derivation of this and also the following results is found in appendix \ref{AHAtransition}. We may now average over the remaining state using the formulae of appendix \ref{Aaveragematrixelements}
\begin{equation} \label{4HAFullResult}
\HA_{\psi,\chi}\left[\abs{\braket{\chi|A|\psi}}^2 \delta (\braket{\chi | \psi} - z)\right] = \frac{N-\abs{z}^2}{N^3-N}\Tr\ AA^{\dag} + \frac{N\abs{z}^2-1}{N^3-N}\abs{\Tr\ A}^2.
\end{equation}
While the former results are valid for uniformly distributed $\ket{\psi}$, we are also interested in the nonuniform case according to section \ref{section2}. To this end we use
\begin{equation}
\ket{\psi} \rightarrow \Lambda \ket{\psi}.
\end{equation}
In general $\Lambda \ket{\psi}$ is not normalized. Thus, we normalize the argument of $\delta \left( \braket{\chi|\Lambda|\psi} - z \right)$ by $\norm{\Lambda \ket{\chi}} = \sqrt{\braket{\chi|\Lambda^2|\chi}}$. This is possible due to the Hermiticity of $\Lambda$. Furthermore as the normalization is not dependent on $\ket{\psi}$, we use the scaling property of the $\delta$-function
\begin{equation}
\delta \left( \frac{\braket{\chi|\Lambda|\psi}}{\sqrt{\braket{\chi|\Lambda^2|\chi}}} - z\right) = \sqrt{\braket{\chi|\Lambda^2|\chi}}\ \delta \left( \braket{\chi|\Lambda|\psi} - z \sqrt{\braket{\chi|\Lambda^2|\chi}} \right).
\end{equation}
In this way we extracted the $\Lambda$-dependence of the absolute value of $\braket{\chi|\Lambda|\psi}$ and are able to retain our choice $\abs{z} \leq 1$ by rescaling. The square root prefactor is cancelled by the normalization, for details see appendix \ref{AHAtransition}. Therefore we arrive at an integral which is equivalent to \eqref{4HAFixedResult} for nonuniformly distributed $\ket{\psi}$
\begin{equation} \label{4HAFixedNonuniformResult}
\HA_{\psi}\left[\abs{\braket{\chi |A\Lambda| \psi}}^2 \delta \left(\frac{\braket{\chi |\Lambda| \psi}}{\sqrt{\braket{\chi|\Lambda^2|\chi}}} - z\right)\right] = \frac{1 - \abs{z}^2}{N - 1} \braket{\chi|A \Lambda^2 A^\dag|\chi} + \frac{N\abs{z}^2 - 1}{N - 1} \frac{\abs{\braket{\chi|A \Lambda^2|\chi}}^2}{\braket{\chi|\Lambda^2|\chi}}.
\end{equation}
It turns out that a subsequent averaging over $\ket{\chi}$ is more difficult than in the uniform case. This is because of the fraction in the second term of \eqref{4HAFixedNonuniformResult}. We treat it with an approximation at the very end of the next section (cf. \eqref{5HAFullApproximation}), where we demonstrate our results numerically. For this purpose we will also calculate the variances of the here 
presented cases.

\section{Numerical demonstration and related aspects} \label{section5}

Our numerical demonstration is based upon the KIC model introduced in section \ref{section3}. The transition probability for individual, randomly sampled $\ket{\chi}$ and $\ket{\psi}$ after action of the KIC Floquet operator \eqref{3Floquet} is shown. The unitarity $UU^\dag = U^\dag U = \unit$ of the Floquet operator leads to slight simplifications of \eqref{4HAFixedResult}, \eqref{4HAFullResult} and \eqref{4HAFixedNonuniformResult}. We will first deal with states, that are uniformly distributed under the imposed conditions. Subsequently, we prepare our states such that they feature a given expectation value of the magnetization \eqref{2Magnetization}, which yields a nonuniform distribution.

\subsection{Uniformly distributed states}

In the following we present a short protocol on how we find the states, that fulfill $z = \braket{\chi|\psi}$, but else are random. First, we introduce the angle between the states $\theta = \Arccos\abs{z}$ \cite{Bengtsson,Scharnhorst}, which we will use to parametrize our numerics. Then we sample uniformly distributed $\ket{\xi}$, so that
\begin{equation} \label{5ChiPerp}
\ket{\chi}_{\perp} = \frac{\ket{\xi} - \braket{\chi|\xi} \ket{\chi}}{\sqrt{1 - \abs{\braket{\chi|\xi}}^2}}
\end{equation}
is normalized and uniformly distributed in the space perpendicular to $\ket{\chi}$. Thus $\ket{\chi}_{\perp}$, unlike $\ket{\chi}$, is not fixed and $\ket{\psi}$ is in dependence of $\theta$ resp. $\abs{z}$
\begin{align} \label{5Psi}
\ket{\psi} & = \cos \theta \ket{\chi} + \sin \theta \ket{\chi}_{\perp} 
\\
& = \abs{z} \ket{\chi} + \sqrt{1 - \abs{z}^2} \ket{\chi}_{\perp}. \nonumber
\end{align}
In this section we plot the analytical results for the Hilbert space average. The individual transition probabilities, displayed as points, will naturally spread around these curves. The width of this spreading is the standard deviation, the square root of the Hilbert space variance. The variance corresponding to \eqref{4HAFixedResult} is
\begin{align} \label{5HVFixedResult}
\HV_{\psi}\left[\abs{\braket{\chi |A| \psi}}^2 \delta (\braket{\chi | \psi} - z)\right] &=\HA_{\psi}\left[\abs{\braket{\chi |A| \psi}}^4 \delta (\braket{\chi | \psi} - z)\right] - \HA_{\psi}\left[\abs{\braket{\chi |A| \psi}}^2 \delta (\braket{\chi | \psi} - z)\right]^2 \nonumber
\\& = \lambda_1\left(N,\abs{z}\right) \braket{\chi |A A^\dag| \chi}^2
 + \lambda_2\left(N,\abs{z}\right) \braket{\chi |A A^\dag| \chi} \abs{\braket{\chi |A| \chi}}^2 
\\ &+ \lambda_3\left(N,\abs{z}\right) \abs{\braket{\chi |A| \chi}}^4. \nonumber 
\end{align}
The second moment of the transition probabilities are obtained in a similar way to the average. This is also addressed in appendix \ref{AHAtransition}. The prefactors
\begin{align} \label{5HVFixedFunctions}
\lambda_1\left(N,\abs{z}\right) & =  \frac{2 \left( 1 - \abs{z}^2 \right)^2}{N (N-1)} - \frac{\left(1  - \abs{z}^2 \right)^2}{(N-1)^2}
\\
\lambda_2\left(N,\abs{z}\right) & = \frac{2 \left( 1 - \abs{z}^2 \right)^2}{(N-1)^2} + \frac{2\left(1 - \abs{z}^2\right)\abs{z}^2}{N - 1} - \frac{4\left(1 - \abs{z}^2\right)^2}{N(N - 1)} \nonumber 
\\
\lambda_3\left(N,\abs{z}\right)) & = \frac{2 \left( 1 - \abs{z}^2 \right)^2}{N (N-1)} - \frac{\left(1 - \abs{z}^2 \right)^2}{(N-1)^2} - \frac{2\left(1 - \abs{z}^2\right) \abs{z}^2}{N - 1} \nonumber
\end{align}
are rational functions in $N$ and $\abs{z}$. It is easy to check that the special cases give the correct results. Choosing $A=\unit$ yields
\begin{align} \label{5HAFixedSpecialCase1}
\HA_{\psi}\left[\abs{\braket{\chi | \psi}}^2 \delta (\braket{\chi | \psi} - z)\right] & = \abs{z}^2
\\
\HV_{\psi}\left[\abs{\braket{\chi | \psi}}^2 \delta (\braket{\chi | \psi} - z)\right] & = 0. \nonumber
\end{align}
Thus, the $\delta$-function acts as expected and the average yields the desired overlap of $\ket{\chi}$ with $\ket{\psi}$, whereas the variance vanishes. This also yields the relation
\begin{equation}
\lambda_1\left(N,\abs{z}\right) + \lambda_2\left(N,\abs{z}\right) + \lambda_3\left(N,\abs{z}\right) = 0.
\end{equation}
Furthermore we find
\begin{align} \label{5HAFixedSpecialCase}
\HA_{\psi}\left[\abs{\braket{\chi |A| \psi}}^2 \delta (\braket{\chi | \psi} - 1)\right] & = \abs{\braket{\chi|A|\chi}}^2
\\
\HV_{\psi}\left[\abs{\braket{\chi |A| \psi}}^2 \delta (\braket{\chi | \psi} - 1)\right] & = 0, \nonumber
\end{align}
because by setting $z=1$ (or any other complex number with unit magnitude) we fix $\ket{\psi}$ along $\ket{\chi}$. The set of states parallel to $\ket{\chi}$ contains only one element (that being $\ket{\chi}$ itself), which explains the variance of zero.

\begin{figure}[t]
\centering
\begin{subfigure}{.5\textwidth}
  \centering
  \includegraphics[width=.9\linewidth]{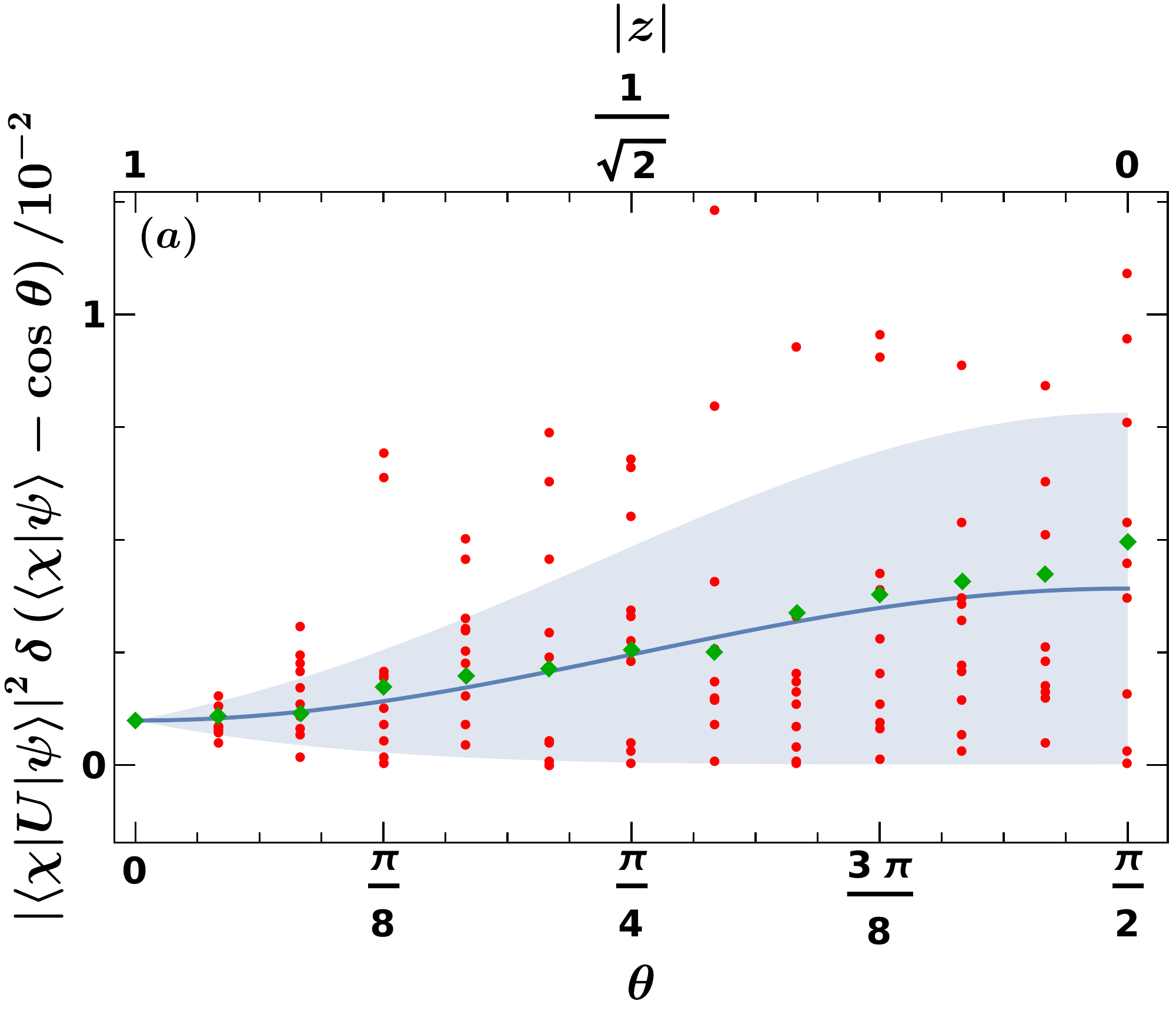}
\end{subfigure}%
\begin{subfigure}{.5\textwidth}
  \centering
  \includegraphics[width=.9\linewidth]{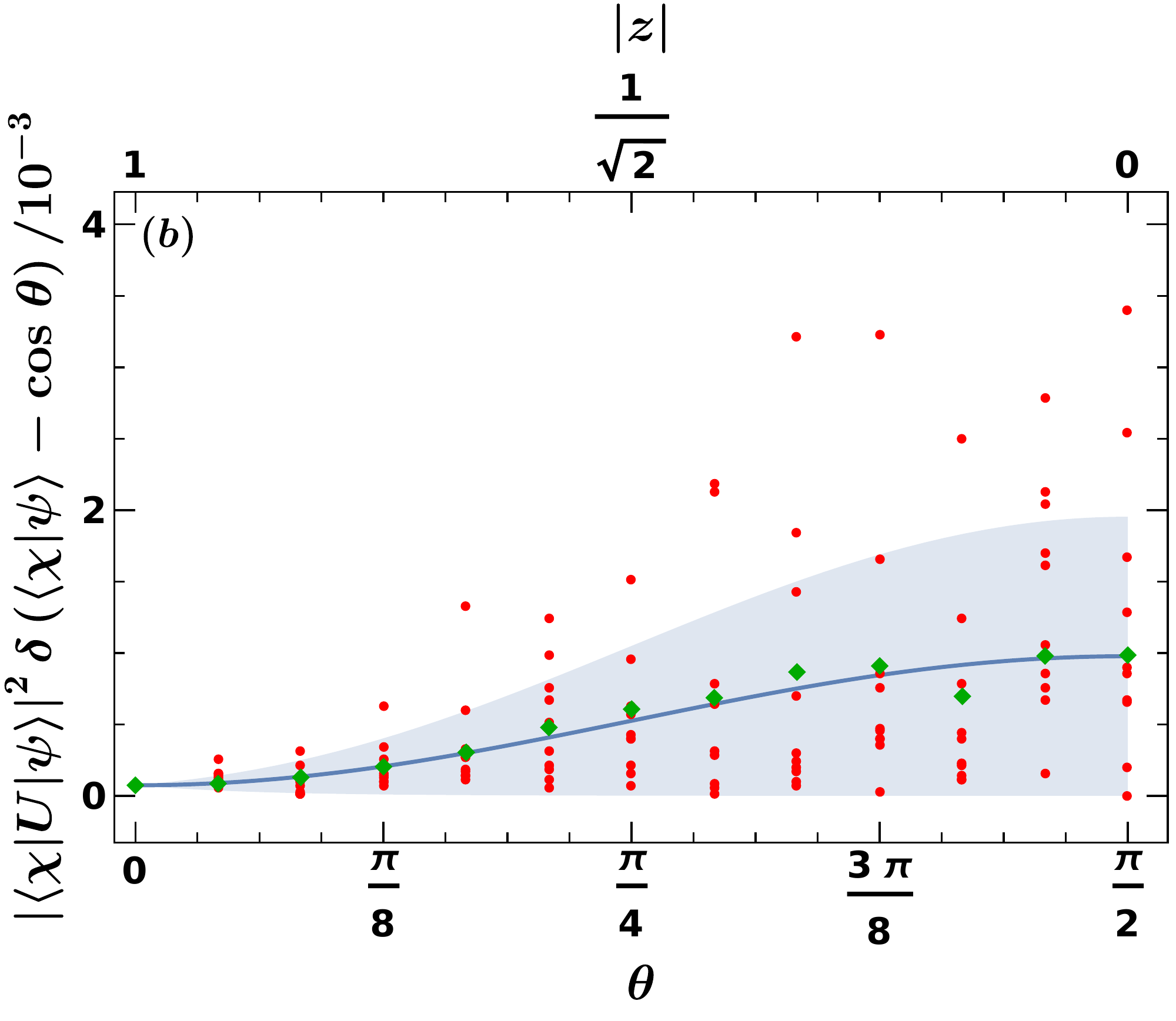}
\end{subfigure}%
\caption{Blue line: Analytical result for the Hilbert space average \eqref{4HAFixedResult}. Red dots: Transition probabilities for 10 individual realizations \eqref{5Psi}. Green diamonds: Their arithmetic mean and the analytical result for the standard deviation according to \eqref{5HVFixedResult} plotted as a tube around the average. The situation is shown for the chaotic chain ($J=b=\pi/4, h=\pi/5$) with the particle numbers (a) $n=8$ and (b) $n=10$. Remember that the map between $\theta$ (lower abscissa) and $\abs{z}$ (upper abscissa) is nonlinear.}
\label{5HAFixed}
\end{figure}

Continuing with an average over both states we arrive at \eqref{4HAFullResult} for the Floquet operator
\begin{align} \label{5HAFullResultFloquet}
\HA_{\psi,\chi}\left[\abs{\braket{\chi|U|\psi}}^2 \delta (\braket{\chi | \psi} - z)\right] & = \frac{N-\abs{z}^2}{N^2-1} + \frac{N\abs{z}^2-1}{N^3-N}\abs{\Tr\ U}^2  
\\
& = \frac{N-\abs{z}^2}{N^2 - 1} + \frac{N\abs{z}^2 -1}{N^2 -1}K(1), \nonumber
\end{align}
where the spectral form factor $K(1) = \abs{\Tr\ U}^2 / N$ emerges. An extension of \eqref{5HAFullResultFloquet} to $U^T$ with arbitrary $T$ is straightforward. The Hilbert space variance
\begin{small}
\begin{equation} \label{5HVFullResultFloquet}
\HV_{\psi,\chi}\left[\abs{\braket{\chi |A| \psi}}^2 \delta (\braket{\chi | \psi} - z)\right] = \HA_{\psi,\chi}\left[\abs{\braket{\chi |A| \psi}}^4 \delta (\braket{\chi | \psi} - z)\right] - \HA_{\psi,\chi}\left[\abs{\braket{\chi |A| \psi}}^2 \delta (\braket{\chi | \psi} - z)\right]^2
\end{equation}
\end{small}
is quite lengthy, which is why we restrict ourselves to stating its individual components directly for the case of the Floquet operator. Exploiting its unitarity we obtain for the second moment
\begin{align} \label{52ndmoment}
\HA_{\psi,\chi}\left[\abs{\braket{\chi |U| \psi}}^4 \delta (\braket{\chi | \psi} - z)\right] &= \frac{2\left(1-\abs{z}^2 \right)^2}{N(N-1)} 
\\
& + \left[ \frac{2 \left( 1- \abs{z}^2\right)^2}{N(N-1)} - \frac{4\left(1 - \abs{z}^2 \right) \abs{z}^2}{N - 1} + \abs{z}^4 \right] \HA_{\chi}\left[\abs{\braket{\chi|U|\chi}}^4\right] \nonumber
\\
&+ \left[\frac{4\left(1 - \abs{z}^2 \right) \abs{z}^2}{N - 1} - \frac{4 \left( 1- \abs{z}^2\right)^2}{N(N-1)}\right] \HA_{\chi}\left[\abs{\braket{\chi|U|\chi}}^2 \right] \nonumber
\end{align}
with
\begin{small}
\begin{align} \label{511}
\HA_{\chi} \left[\abs{\braket{\chi |U| \chi}}^2 \right] &= \frac{1}{N(N+1)} \left( \abs{\Tr\  U}^2 + N \right)
\\
\HA_{\chi} \left[\abs{\braket{\chi |U| \chi}}^4 \right] & = \frac{(N-1)!}{(N+3)!} \left( \abs{\Tr\ U}^4 + 2\ \text{Re}\left( \Tr\ U^2\ \Tr^2\ U^\dag \right) + \abs{\Tr\ U^2}^2 + (4N + 8) \abs{\Tr\ U}^2 + 2N^2 + 6N \right). \nonumber
\end{align}
\end{small}
The second term in the Hilbert space variance \eqref{5HVFullResultFloquet} is given by the square of \eqref{5HAFullResultFloquet}
\begin{align} \label{5HAsquared}
\HA_{\psi,\chi}\left[\abs{\braket{\chi|U|\psi}}^2 \delta (\braket{\chi | \psi} - z)\right]^2 &= \left( \frac{N - \abs{z}^2}{N^2 - 1} \right)^2 + \left( \frac{N \abs{z}^2 - 1}{N^3 - N} \right)^2 \abs{\Tr\ U}^4 
\\
&+ 2 \left(\frac{N - \abs{z}^2}{N^2 - 1}\right)\left( \frac{N\abs{z}^2 -1}{N^3-N}\right) \abs{\Tr\ U}^2. \nonumber
\end{align}
We observe the interesting feature that the Hilbert space average of the transition probabilities are fully determined by the spectral statistics of the deployed operator. However, this is not possible for the variance, because of the term $\text{Re}\left( \Tr\  U^2\ \Tr^2\ U^\dag \right)$ occurring in $\HA_{\chi} \left[\abs{\braket{\chi |U| \chi}}^4 \right]$.

In Fig. \ref{5HAFixed} we display the transition probability for a randomly chosen, but fixed state $\ket{\chi}$, corresponding to \eqref{4HAFixedResult}. We use chaotic system parameters ($J=b=\pi/4, h=\pi/5$) for our KIC Floquet operator $U$. The most notable thing one can infer is the increase in variance with growing angle $\theta$, starting from the point $\theta = 0$, where it vanishes. We note that at this point the Hilbert space average is solely depending on $\ket{\chi}$ (see \eqref{4HAFixedResult}). However, an implicit $N$-dependence emerges as it becomes clear from the 
expression obtained after averaging over the remaining state, see \eqref{5HAFullResultFloquet}. At the value $\theta=\pi/2$ the variance, as well as the measure of accessible states $\ket{\psi}$, that fulfill $\braket{\chi|\psi} = z$, are maximized.

Furthermore, in Fig.\ \ref{5HAFixed} we can observe that the Hilbert space variance is not negligibly small compared to the Hilbert space average. A sharp Hilbert space average, like in the case of the expectation values, discussed in section \ref{section2}, can be achieved by demanding traces of functions of $U$ (see \eqref{511}) to be of order $N$. In that way we require $\abs{\Tr\ U}^2$ to be of the order $N^2$ to reach a constant Hilbert space average in (5.8) and obtain similarly a variance of the order $1/N$. However, it turns out that this assumption for the traces of $U$ is not fulfilled for our specific $U$: in \cite{Akila} it was shown that $\abs{\Tr\ U}^2$ is for our $U$ of the order $N^c$ with $c\leq1$ in a wide parameter range. In fact, for our specific $U$ the Hilbert space average can be sharp only in a limited amount of special cases. Two of these cases were discussed in \eqref{5HAFixedSpecialCase1} and \eqref{5HAFixedSpecialCase}. The third option is that $\ket{\chi}$ is an eigenstate of the deployed operator, which has the same effect as $U=\unit$ in equations \eqref{4HAFixedResult} and \eqref{5HVFixedResult}.

\begin{figure}[t]
\centering
\begin{subfigure}{.5\textwidth}
  \centering
  \includegraphics[width=.9\linewidth]{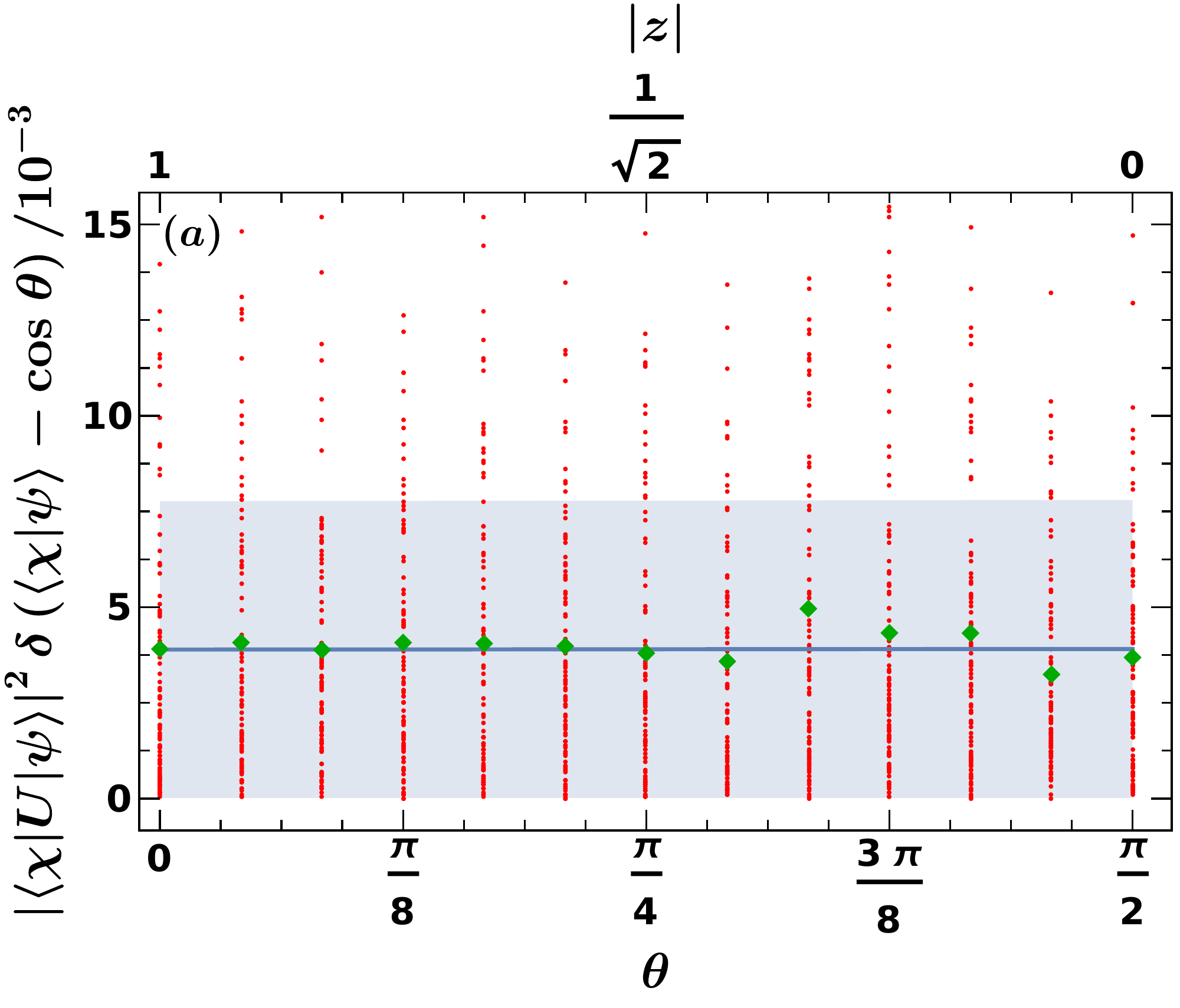}
\end{subfigure}%
\begin{subfigure}{.5\textwidth}
  \centering
  \includegraphics[width=.9\linewidth]{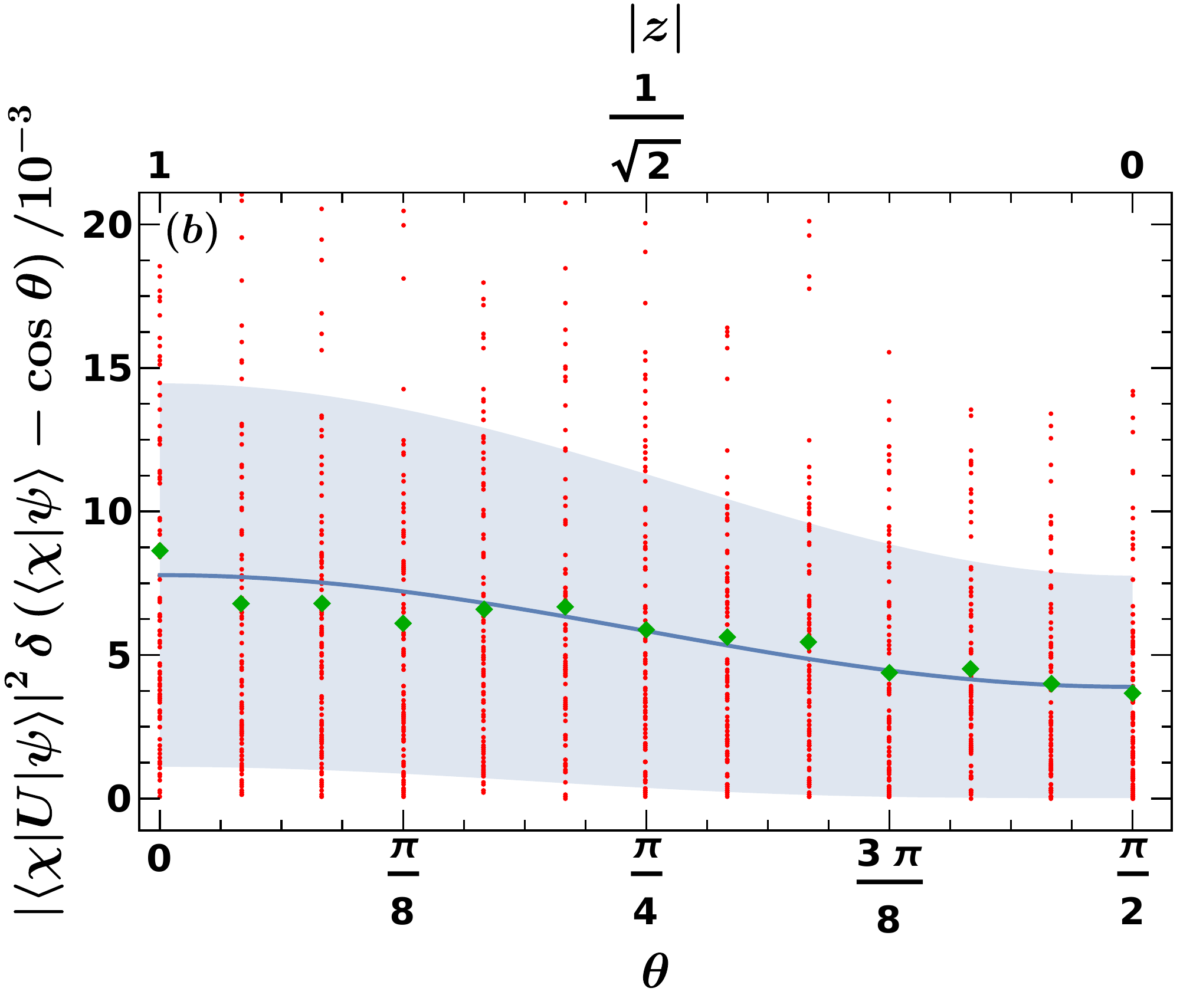}
\end{subfigure}%
\caption{Blue line: Analytical result for the Hilbert space average \eqref{5HAFullResultFloquet}. Red dots: Transition probabilities for 100 individual realizations. Green diamonds: Their arithmetic mean and the analytical result for the standard deviation according to \eqref{5HVFullResultFloquet} plotted as a tube around the average. We show the situation for an $n=8$-particle spin chain, that is (a) chaotic ($J=b=\pi/4, h=\pi/5$) and (b) interactionless ($J=h=0, b=\pi/4$).}
\label{5HAFull}
\end{figure}

In Fig.\ \ref{5HAFull} the Hilbert space average \eqref{5HAFullResultFloquet} is shown for a completely chaotic system ($J=b=\pi/4$, $h=\pi/5$) on the left and for an integrable system ($J=h=0, b=\pi/4$) on the right. As opposed to the case, in which we kept one state fixed (see Fig.\ \ref{5HAFixed}), we do not observe a large dynamics of the transition probability in dependence of $\theta$. The derivative of \eqref{5HAFullResultFloquet}
\begin{equation} \label{5SFFSlopeFull}
\partial_{\abs{z}} \HA_{\psi,\chi}\left[\abs{\braket{\chi |U| \psi}}^2 \delta (\braket{\chi | \psi} - z)\right] = \frac{2\abs{z}}{N^2 - 1} \left( N K(1) - 1 \right)
\end{equation}
determines this behaviour. We can infer, that for $K(1) \gg 1/N$ the probability should grow with $\abs{z}$ and decrease for $K(1) \ll 1/N$. However, the prefactor $1/\left( N^2 - 1 \right)$ ensures that the derivative is practically zero for large Hilbert space dimensions as can be seen for the chaotic case, shown in the left panel of Fig.\ \ref{5HAFull}. In the integrable case we chose the Floquet operator as $U = \bigotimes_{i=1}^n \Exp\left[-i \frac{\pi}{4} \sigma^x_i \right]$, i.e. we rotate every spin by $\pi/2$ around the $x$-axis, while there is no interaction between the spins. Here a large value for the spectral form factor, $K(T) = 1$, is found for all times by straightforward, exact calculation. And indeed we find a positive slope in $\abs{z}$, as the expression on the right hand side of \eqref{5SFFSlopeFull} is one order larger in $N$ than for $K(T) \ll 1/N$.

\begin{figure}[t]
\centering
\begin{subfigure}{.5\textwidth}
  \centering
  \includegraphics[width=.9\linewidth]{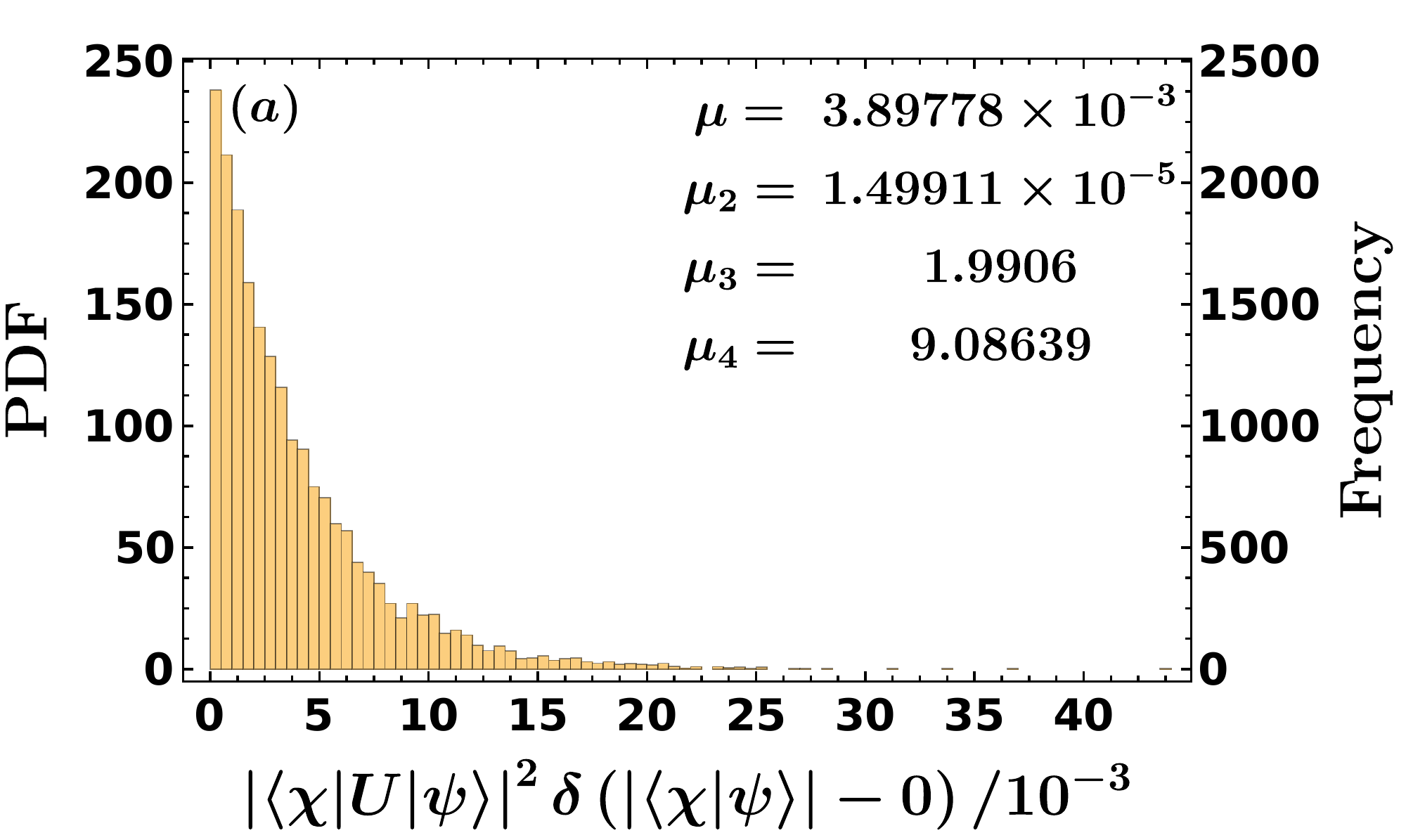}
\end{subfigure}%
\begin{subfigure}{.5\textwidth}
  \centering
  \includegraphics[width=.9\linewidth]{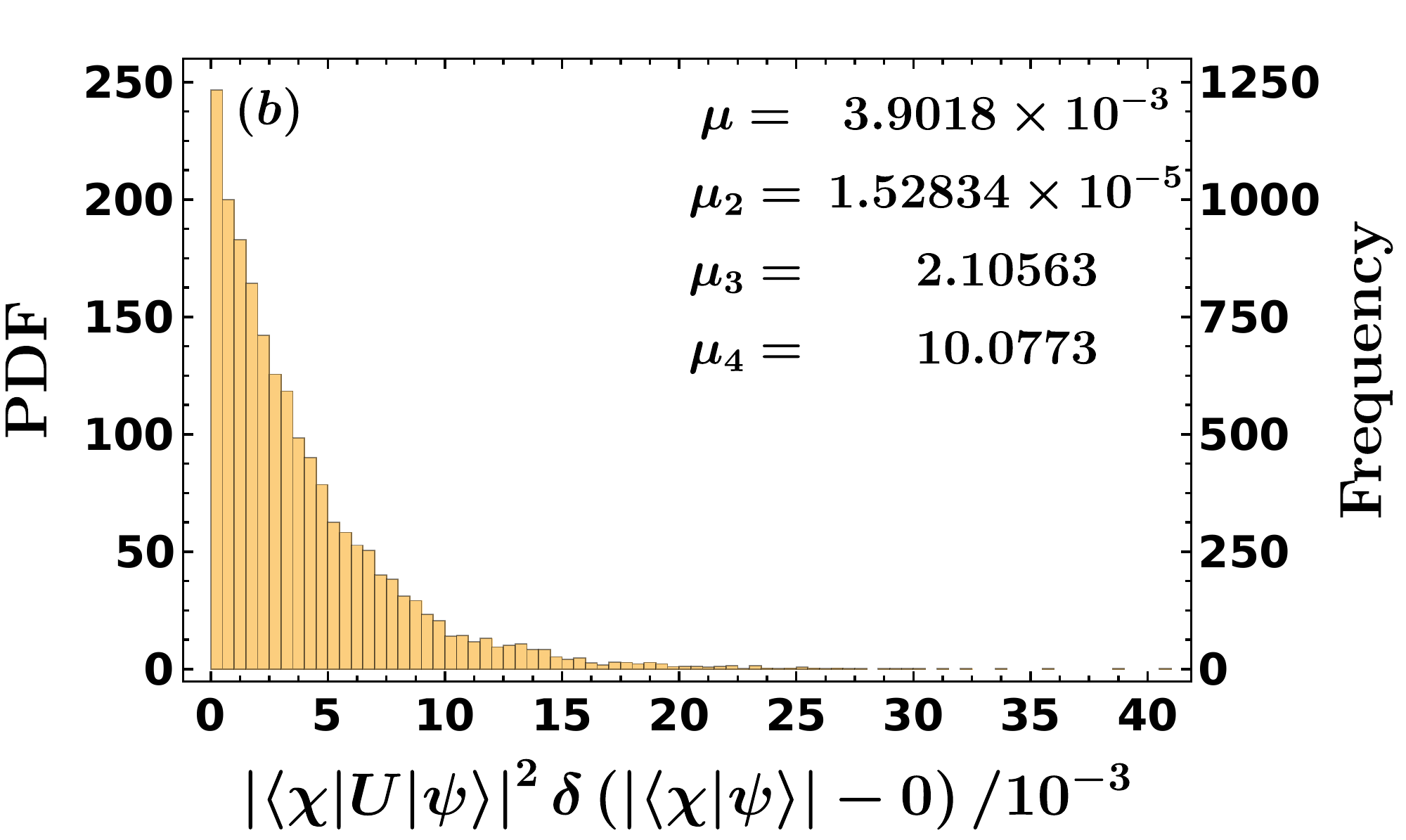}
\end{subfigure}%
\caption{Histogram of transition probabilities for the Floquet operator $U$ of a kicked Ising chain (a) within the chaotic regime ($J=b=\pi/4, h=\pi/5$) and (b), that is interactionless ($J=h=0, b=\pi/4$). The inset contains the mean $\mu$, variance $\mu_2$, skewness $\mu_3$ and kurtosis $\mu_4$.}
\label{4HAFullHisto}
\end{figure}

Furthermore we study the distribution of transition probabilities $\abs{\braket{\chi|U|\psi}}^2$ under the secondary condition $\braket{\chi|\psi} = z$. Without this condition one finds a distribution of the Kumaraswamy type
\begin{equation} \label{5Dist}
p(s) = \left( N-1 \right) \left( 1 - s \right)^{N-2},
\end{equation}
which replaces the Porter-Thomas distribution \cite{Alhassid2} in the case of broken time reversal invariance. We present a detailed derivation of this result in appendix \ref{ADist}. It converges to an exponential distribution $Ne^{-Ns}$ for large $N$. Hence, the latter distribution and \eqref{5Dist} share the same properties of equal mean and standard deviation
\begin{equation}
\mu = \sqrt{\mu_2} = \frac{1}{N}
\end{equation}
and parameter independent skewness and kurtosis
\begin{equation}
\mu_3 = 2  \qquad  \qquad \mu_4 = 9
\end{equation}
in this limit. In Fig.\ \ref{4HAFullHisto} we show histograms of the transition probability $\abs{\braket{\chi|U|\psi}}^2$ for 10000 individual realizations of $\ket{\chi}$ and $\ket{\psi}$ and under the condition $\abs{\braket{\chi|\psi}} = 0$, i.e. $\theta = \pi/2$. Once more the left side is obtained from a chaotic Floquet operator ($J=b=\pi/4$, $h=\pi/5$) and the right side from the one, that shows no interaction and only rotates every spin ($J=h=0, b=\pi/4$). Our numerically obtained distributions are convincingly close to \eqref{5Dist}. This is due to the assumption of statistical orthogonality entering in the derivation of \eqref{5Dist}, which is well fulfilled in a high dimensional space. For different values of $z$ we observe also an exponential distribution, but with deviating parameters $\mu$, $\mu_2$, $\mu_3$ and $\mu_4$. Thus, apart from the trivial case $U = \unit$ (yielding the distribution $\delta ( s - \abs{z}^2 )$), the information about the initial $z$ is erased by the time evolution.

\subsection{Nonuniformly distributed states}

We now turn our attention to the situation in which the states are not uniformly distributed. For the moment we use in our formulae the operator $A$ again, instead of unitary $U$, in order to emphasize that they are valid for arbitrary operators. Keeping one of the states fixed the Hilbert space average is given by \eqref{4HAFixedNonuniformResult}. The variance is
\begin{align}\label{5HVFixedNonuniformResult}
\HV_\psi \left[\abs{\braket{\chi |A\Lambda| \psi}}^2 \delta \left(\frac{\braket{\chi |\Lambda| \psi}}{\sqrt{\braket{\chi|\Lambda^2|\chi}}} - z\right)\right] & = \lambda_1\left(N,\abs{z}\right) \braket{\chi |A \Lambda^2 A^\dag| \chi}^2
\\
& + \lambda_2\left(N,\abs{z}\right) \braket{\chi |A \Lambda^2 A^\dag| \chi} \frac{\abs{\braket{\chi |A \Lambda^2| \chi}}^2}{\braket{\chi|\Lambda^2|\chi}} \nonumber
\\
& + \lambda_3\left(N,\abs{z}\right) \frac{\abs{\braket{\chi |A \Lambda^2 | \chi}}^4}{\braket{\chi|\Lambda^2|\chi}^2}, \nonumber
\end{align}
where the prefactors are the same as in \eqref{5HVFixedFunctions}. It is easy to check that \eqref{4HAFixedNonuniformResult} and \eqref{5HVFixedNonuniformResult} are equal to \eqref{4HAFixedResult} and \eqref{5HVFixedResult} for $\Lambda = \unit$.

\begin{figure}[t]
\centering
\begin{subfigure}{.5\textwidth}
  \centering
  \includegraphics[width=.9\linewidth]{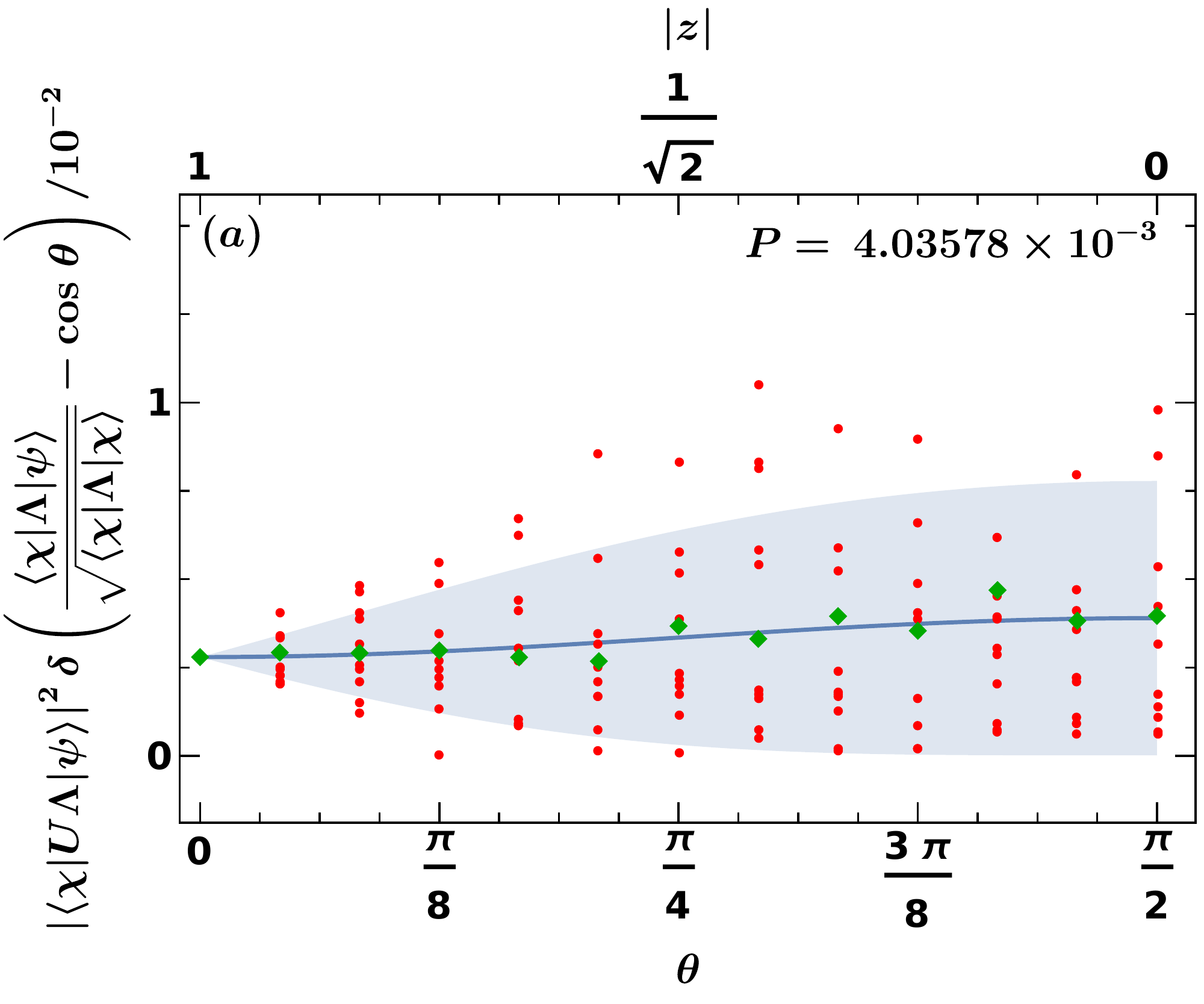}
\end{subfigure}%
\begin{subfigure}{.5\textwidth}
  \centering
  \includegraphics[width=.9\linewidth]{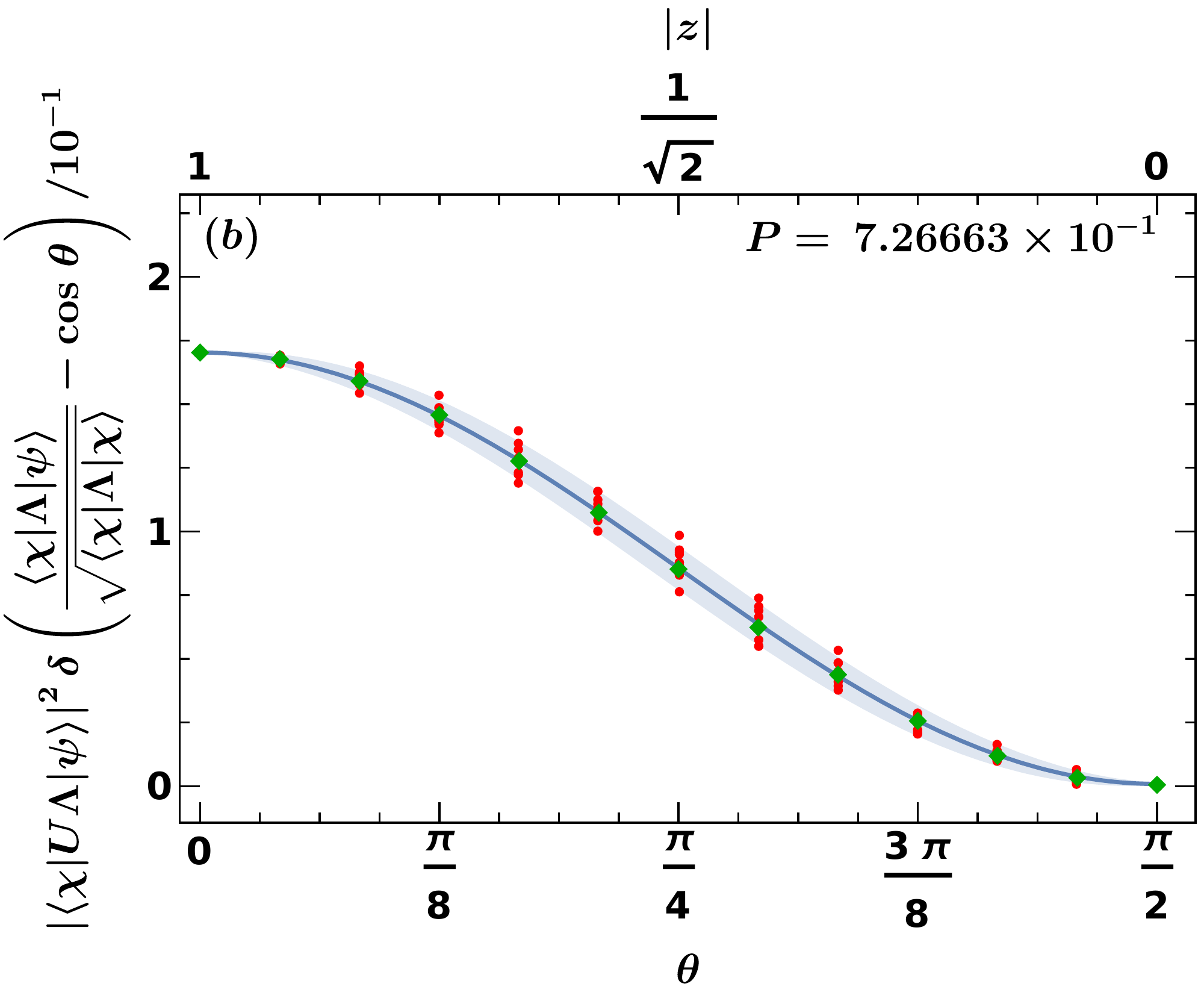}
\end{subfigure}
\caption{Blue line: Analytical result for the Hilbert space average \eqref{4HAFixedNonuniformResult}. Red dots: Transition probabilities for 10 individual realizations \eqref{5LPsi}. Green diamonds: Their arithmetic mean and the analytical result for the standard deviation according to \eqref{5HVFixedNonuniformResult} plotted as a tube around the average. The inset shows the purities of the deployed statistical operators, which are determined by the magnetization $m_z$. It is $m_z = 0.5$ for (a) and $m_z = 7$ for (b).}
\label{5HAFixedNonuniform}
\end{figure}

Now we want to average over the remaining state $\ket{\chi}$ in \eqref{4HAFixedNonuniformResult}. By setting $\ket{\chi} \rightarrow \Lambda' \ket{\chi}$ we take into account that it may as well feature a nonuniform distribution. The average of the first term in \eqref{4HAFixedNonuniformResult}
\begin{equation}
\HA_{\chi} \left[\braket{\chi|\Lambda' A \Lambda^2 A^\dag \Lambda'|\chi}\right] = \frac{\Tr\ \Lambda' A \Lambda^2 A^\dag \Lambda'}{N}
\end{equation}
is known. However, an exact analytical calculation of Hilbert space averages of the kind \linebreak $\HA_{\chi} \left[\abs{\braket{\chi|\alpha|\chi}}^2 / \braket{\chi|\beta|\chi}\right]$ with arbitrary matrices $\alpha$ and $\beta$, which is needed for the second term in \eqref{4HAFixedNonuniformResult}, is beyond the scope of this paper. Therefore we are using the geometric series $1/(1-x) = 1 + x + x^2 + \mathcal{O}\left( x^3 \right)$ as an approximation for $\abs{x} \ll 1$. This yields
\begin{align}
\frac{1}{\braket{\chi|\Lambda' \Lambda^2 \Lambda'|\chi}} & = \frac{1}{1 - \left( 1 - \braket{\chi|\Lambda' \Lambda^2 \Lambda'|\chi} \right)}
\\
& = 1 + \left( 1 - \braket{\chi|\Lambda' \Lambda^2 \Lambda'|\chi} \right) + \left( 1 - \braket{\chi|\Lambda' \Lambda^2 \Lambda'|\chi} \right)^2 + \mathcal{O}\left( \left( 1 - \braket{\chi|\Lambda' \Lambda^2 \Lambda'|\chi} \right)^3 \right) \nonumber
\\
& = 3 - 3 \braket{\chi|\Lambda' \Lambda^2 \Lambda'|\chi} + \braket{\chi|\Lambda' \Lambda^2 \Lambda'|\chi}^2 + \mathcal{O}\left( \left( 1 - \braket{\chi|\Lambda' \Lambda^2 \Lambda'|\chi} \right)^3 \right). \nonumber
\end{align}
Thus, we can write the approximated Hilbert space average as
\begin{align} \label{5HAFullApproximation}
\HA_{\chi} \left[ \frac{\abs{\braket{\chi|\Lambda' A \Lambda^2 \Lambda'|\chi}}^2}{\braket{\chi|\Lambda' \Lambda^2 \Lambda'|\chi}}\right] \approx 3\ &\HA_{\chi} \left[ \abs{\braket{\chi|\Lambda' A \Lambda^2 \Lambda'|\chi}}^2 \right] - 3\ \HA_{\chi} \left[ \braket{\chi|\Lambda' \Lambda^2 \Lambda'|\chi} \abs{\braket{\chi|\Lambda' A \Lambda^2 \Lambda'|\chi}}^2 \right] \nonumber
\\
 + &\HA_{\chi} \left[ \braket{\chi|\Lambda' \Lambda^2 \Lambda'|\chi}^2 \abs{\braket{\chi|\Lambda' A \Lambda^2 \Lambda'|\chi}}^2 \right],
\end{align}
where we deploy formula \eqref{AB1toBM} of appendix \ref{Aaveragematrixelements} to calculate the occurring Hilbert space averages. The variance is
\begin{align} \label{5HVFullApproximation}
\HV_{\psi,\chi}\left[\abs{\braket{\chi |\Lambda' A \Lambda| \psi}}^2 \delta \left(\frac{\braket{\chi |\Lambda' \Lambda| \psi}}{\sqrt{\braket{\chi |\Lambda' \Lambda^2 \Lambda'| \chi}}} - z\right)\right] & =  \HA_{\psi,\chi}\left[\abs{\braket{\chi|\Lambda' A \Lambda|\psi}}^4\delta \left(\frac{\braket{\chi |\Lambda' \Lambda| \psi}}{\sqrt{\braket{\chi |\Lambda' \Lambda^2 \Lambda'| \chi}}} - z\right) \right] 
\\
&- \HA_{\psi,\chi}\left[\abs{\braket{\chi|\Lambda' A \Lambda|\psi}}^2 \delta \left(\frac{\braket{\chi |\Lambda' \Lambda| \psi}}{\sqrt{\braket{\chi |\Lambda' \Lambda^2 \Lambda'| \chi}}} - z\right) \right]^2, \nonumber
\end{align}
which contains $\HA_{\chi} \left[ \abs{\braket{\chi|\alpha|\chi}}^2/\braket{\chi|\beta\chi} \right]$ and $\HA_{\chi} \left[ \abs{\braket{\chi|\alpha|\chi}}^4/\braket{\chi|\beta\chi}^2 \right]$. Also for the latter expression we would again be able to approximate the fraction via Taylor expansion and then apply \eqref{AB1toBM}, which leads to a fully analytical approximation of the variance. However, as this procedure turns out to be quite cumbersome, we will settle with a numerical calculation of the Hilbert space variance, when testing our approximation of the average.

For the purpose of a numerical illustration we start with \eqref{4HAFixedNonuniformResult} and \eqref{5HVFixedNonuniformResult}, where one of the states is still fixed. As operator $A$ we use again the KIC Floquet operator $U$. Furthermore we will use $\Lambda$ to center the states $\ket{\psi}$ around a preset expectation value $m_z$ of the magnetization in $z$-direction
\begin{equation}
M_z = \sum_{i=1}^n \sigma_i^z.
\end{equation}
The expectation value of the uniform ensemble, ergo the completely mixed state $\rho_{\text{m}}$, is $\Tr\ M_z \rho_{\text{m}} = 0$. We choose now instead $\rho_{m_z}$, defined by $\Tr\ M_z \rho_{m_z} = m_z$ with the operator \eqref{2RhoReimann}. Naturally, $m_z$ has to lie in the measurement range $[-n,n]$ of $M_z$.

\begin{figure}[t]
\centering
\begin{subfigure}{.5\textwidth}
  \centering
  \includegraphics[width=.9\linewidth]{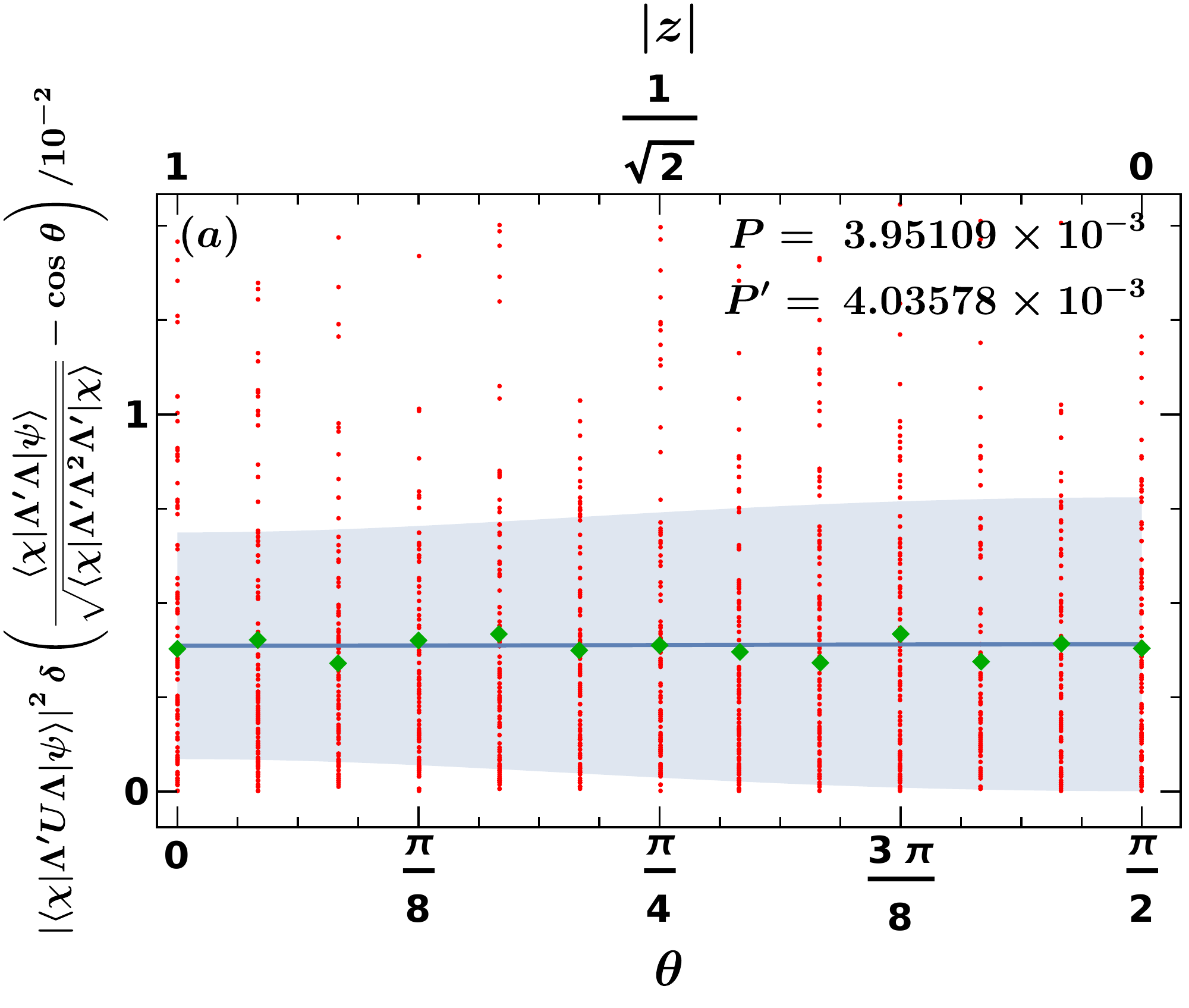}
\end{subfigure}%
\begin{subfigure}{.5\textwidth}
  \centering
	\includegraphics[width=.9\linewidth]{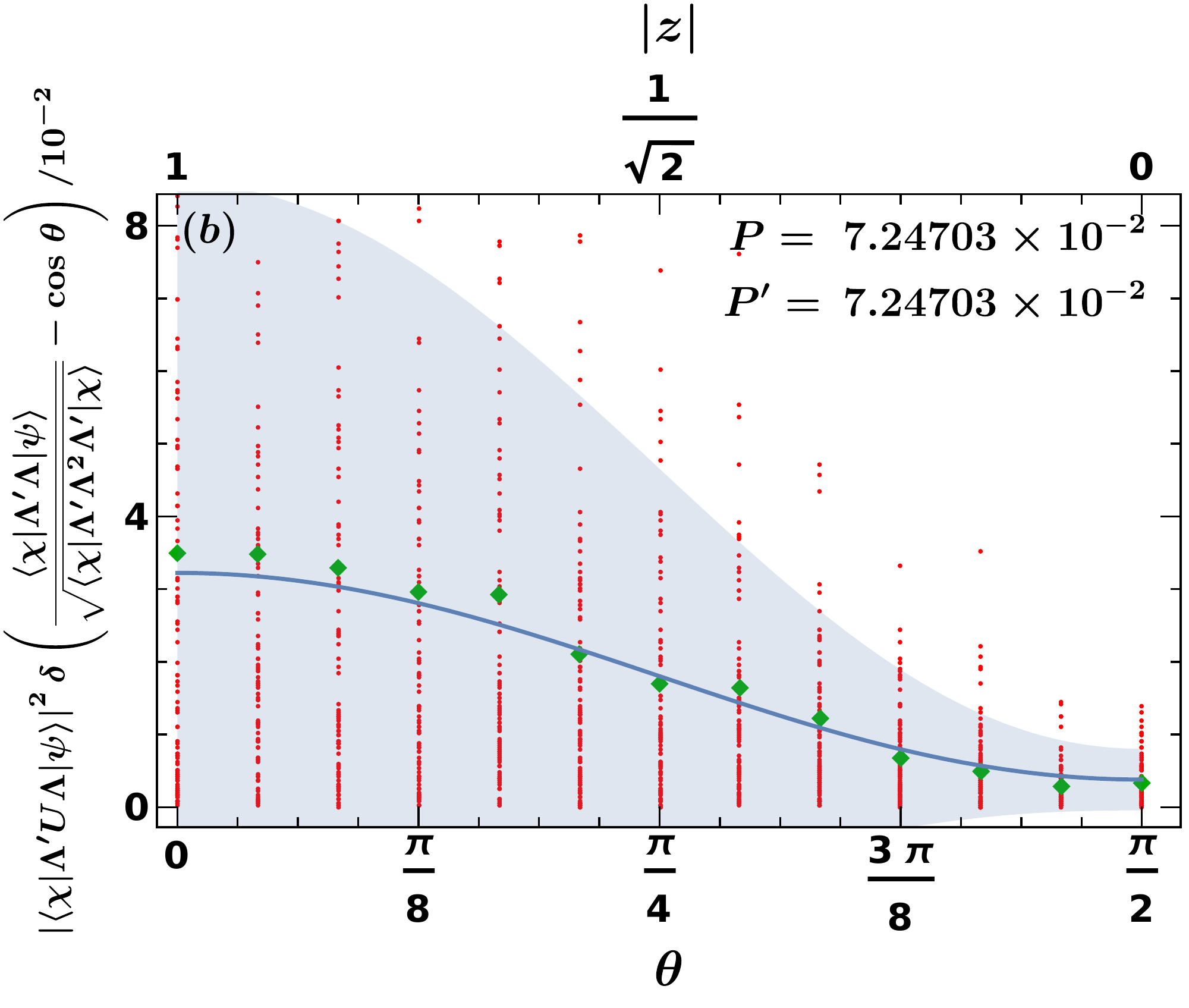}
\end{subfigure}
\caption{Blue line: Analytical result for the Hilbert space average using the approximation \eqref{5HAFullApproximation}. Red dots: Transition probabilities for 100 individual realizations \eqref{5LPsi}. Green diamonds: Their arithmetic mean and the analytical result for the standard deviation according to \eqref{5HVFullApproximation} plotted as a tube around the average. The inset shows the purities of the deployed statistical operators $\rho$ and $\rho'$. They are determined by the magnetizations $m_z=0.5$, $m_z' = -0.3$ for (a) and $m_z=3$,$m_z'=-3$ for (b).}
\label{5HAFullApproximationFig}
\end{figure}

We find the states $\ket{\psi}$, that fulfill $\braket{\chi|\Lambda|\psi} = z \sqrt{\braket{\chi|\Lambda^2|\chi}}$, but else are uniformly distributed, by using $\ket{\chi}_\perp$ from \eqref{5ChiPerp}
\begin{align} \label{5LPsi}
\ket{\psi} & = \cos \theta \frac{\Lambda \ket{\chi}}{\norm{\Lambda \ket{\chi}}} + \sin \theta \frac{\Lambda^{-1} \ket{\chi}_{\perp}}{\norm{\Lambda^{-1} \ket{\chi}_{\perp}}}
\\
& = \abs{z} \frac{\Lambda \ket{\chi}}{\norm{\Lambda \ket{\chi}}} + \sqrt{1 - \abs{z}^2} \frac{\Lambda^{-1} \ket{\chi}_{\perp}}{\norm{\Lambda^{-1} \ket{\chi}_{\perp}}}. \nonumber
\end{align}
This procedure is completely analogue to that in \eqref{5Psi}. We want to mention again that while the vector $\Lambda \ket{\chi}/\norm{\Lambda \ket{\chi}}$ is fixed, $\Lambda^{-1} \ket{\chi}_{\perp}/\norm{\Lambda^{-1} \ket{\chi}_{\perp}}$ is not as we draw $\ket{\chi}_{\perp}$ uniformly from the set of all states, that are perpendicular to $\ket{\chi}$.

We still need to discuss the validity of our assumption $\braket{\chi|\Lambda' \Lambda^2 \Lambda'|\chi} \approx 1$. Its average is given by
\begin{equation}
\HA_{\chi} \left[ \braket{\chi|\Lambda' \Lambda^2 \Lambda'|\chi} \right] = \frac{\Tr\ \Lambda' \Lambda^2 \Lambda'}{N} = N\ \Tr\ \rho \rho'.
\end{equation}
Therefore in the case of $\rho = \rho'$ we have $\HA_{\chi} \left[ \braket{\chi|\Lambda^4|\chi} \right] = N P$ and thus $P$ needs to be close to its lower bound, which is $1/N$. Generally we can estimate it by
\begin{equation}
N^2 p_{\text{min}}\ p'_{\text{min}} \leq \HA_{\chi} \left[ \braket{\chi|\Lambda'\Lambda^2\Lambda'|\chi} \right] = N \sum_{i,j=1}^N p_i\ p_j' \abs{\braket{i|j}}^2 \leq N^2 p_{\text{max}}\ p'_{\text{max}},
\end{equation}
where $p_i$ and $p_j'$ are the eigenvalues of $\rho$ and $\rho'$ in their respective eigenbasis. Thus, if both, the maximal eigenvalues $p_{\text{max}}, p'_{\text{max}}$ and the minimal eigenvalues $p_{\text{min}}, p'_{\text{min}}$ of the deployed operators are near $1/N$ a value close to one is guaranteed. The minimal and maximal eigenvalues are also lower resp. upper bounds for the purity. Consequently, the former requirement implies a purity close to the minimum of $1/N$.

In general, the connection between the purity and the preset expectation value $m_z$ is not easy to find. However, in \cite{Reimann1} it was shown, that for the choice \eqref{2RhoReimann}, $p_{\text{max}}$ is monotonically increasing for $m_z>0$ and monotonically decreasing for $m_z<0$. In the following we will assume that magnetization values $m_z$ close to zero evoke low purity statistical operators and the closer we go to the lower or upper bound $\pm n$ the higher the purity will be.

In Fig.\ \ref{5HAFixedNonuniform} the transition probability is shown for a low purity ($m_z=0.5$) on the left side and for a high purity ($m_z=7$) on the right. In both cases we work within the chaotic regime ($J=b=\pi/4$, $h = 2\pi/5$) of an $n=8$-particle KIC. Again we find a good agreement between our analytical results and the numerical data. We observe that for a low purity there is not much difference compared to the results in Fig.\ \ref{5HAFixed}. In the case of high purity we notice two things.  First, there is a drastic change of the transition probability in dependence of $\theta$ and second, the variance does not follow the same scheme as before. Namely, it does not grow larger with increasing $\theta$, but stays small for all angles. This is because $\Lambda \ket{\psi}$ basically consists of eigenvectors, for which the corresponding eigenvalues are centered around the expectation value of the deployed statistical operator. In the present case this is $m_z = 7$, where the eigenvalue density of our specific observable is quite sparse (cf. end of section \ref{section2}). In other words the amount of states that fulfill $\braket{\psi| \Lambda M_z \Lambda |\psi} \approx 7$ and $\braket{\chi|  \Lambda |\psi}/\sqrt{\braket{\chi|  \Lambda^2 |\chi}} = z$ is small compared to those that solely fulfill the latter condition, which leads to the overall small variance.

Next we consider the average with respect to both states and test our approximation \eqref{5HAFullApproximation}. Here we use $\Lambda'$, just like $\Lambda$, to center the states $\ket{\chi}$ around a certain expectation value of $M_z$. In Fig.\ \ref{5HAFullApproximationFig} we present again a realization with low purities ($m_z = 0.5$, $m'_z = -0.3$) and one with higher purities ($m_z = -m'_z = 3$), but still within the range of validity of our approximation. As expected the low purity case does not differ noticeable from its uniform counterpart in Fig.\ \ref{5HAFull}. In the high purity case we observe a similar effect as in Fig.\ \ref{5HAFixedNonuniform}, where we left one state fixed. The variance at $\theta = 0$ is maximized as this condition forces the states to be composed of similar eigenvectors. This happens at the cost of their expectation values concerning $M_z$, which are still centered around the preset values, but feature also a large variance. As the eigenvalue density becomes more dense in the center of the spectrum, there is a large amount of different options to fulfill $\theta = 0$, leading to the large variance of the transition probability. This is exactly the other way around for $\theta = \pi/2$, where the states can be arranged in such way, that both conditions are fulfilled.

\section{Conclusion and Outlook}

Typicality is a statistical pattern, that occurs with immense probability under the requirement of a large Hilbert space in quantum mechanics. Thus, it belongs to the high dimensionality phenomena, that makes it ideally suited for many-body systems, in which the Hilbert space dimension usually grows exponentially in the particle number.

We adopted the methods of quantum typicality \cite{Bartsch, Gemmer, Reimann1}, namely the Hilbert space average, and applied them to transition probabilities, particularly those where the initial overlap between the deployed states is fixed. By doing this we put the statistical orthogonality, commonly used in random matrix theory, aside. We explored the limits of the Hilbert space averaging methods analytically by averaging first over one state and subsequently over the remaining one. While doing so we used different distributions of states, uniform distributions as well as nonuniform ones, where the states initially feature an expectation value of a given observable close to a desired preset value. We compared the resulting equations for the average and the variance with numerical calculations, using the unitary Floquet operator of a kicked spin chain. Furthermore we could connect the transition probability for uniformly distributed states to the spectral statistics of the deployed operator.

We studied also the statistics of the transition probabilities under the additional condition $\abs{\braket{\chi|\psi}} = z$. For $z=0$ we confirmed the Kumaraswamy distribution from random matrix theory in the case of no constraint on $\abs{\braket{\chi|\psi}}$.

More generally our results apply to transition probabilities involving an arbitrary operator $A$. One interesting future application was already addressed in the introduction. In the doorway mechanism \cite{Kohler1, Kohler2} the doorway state, that is distinguished by some special feature (e.g. collective motion in many-body systems), is coupled to a background of energy eigenstates. The doorway state is not an eigenstate of the Hamiltonian, leading to nonorthogonality  with the background. Another application appears in the equilibration of isolated quantum systems. According to the eigenstate thermalization hypothesis \cite{Reimann2,Steinigeweg3, Khodja, Steinigeweg4} this is the case if all diagonal elements of a given observable in the energy eigenbasis center around a common value while the off-diagonal elements vanish. A similar approach has been introduced in \cite{Luck1,Luck2}, where the trace of the matrix of time averaged transition probabilities in such a system contains information about the equilibration properties. Finally we would also consider it interesting to apply the methods developed here to different quantities. One popular example are reduced statistical operators, that emerge after tracing out pure typical states, that allow to compute for example the entanglement entropy between subsystems.

\section{Acknowledgement}

We thank Petr Braun mainly for fruitful discussions and a careful reading of the manuscript.

\newpage
\appendix
\section{Appendix} \label{Appendix}
\subsection{Hilbert space average of transition probabilities} \label{AHAtransition}

In this appendix we will derive \eqref{4HAFixedNonuniformResult} and \eqref{5HVFixedNonuniformResult} of the main text. Equation \eqref{4HAFixedResult} and \eqref{5HVFixedResult} follow by setting $\Lambda = \unit$. So we want to calculate
\begin{small}
\begin{equation} \label{AIntegral}
\HA_{\psi}\left[\abs{\braket{\chi |A\Lambda| \psi}}^2 \delta \left(\frac{\braket{\chi |\Lambda| \psi}}{\sqrt{\braket{\chi|\Lambda^2|\chi}}} - z\right)\right] = \frac{\int d[\psi] \abs{\braket{\chi |A \Lambda| \psi}}^2 \delta \left(\braket{\chi |\Lambda| \psi} - z \sqrt{\braket{\chi|\Lambda^2|\chi}} \right) \delta (\braket{\psi|\psi} -1)}{\int d[\psi] \delta \left(\braket{\chi |\Lambda|\psi} - z \sqrt{\braket{\chi|\Lambda^2|\chi}}\right) \delta (\braket{\psi|\psi} -1)}.
\end{equation}
\end{small}
Our strategy is that we convert the integrals into a multidimensional Gaussian integrals. To this end we write the delta functions as Fourier integrals
\begin{small}
\begin{align} \label{4FourierSubstitutions}
\delta  (\braket{\psi|\psi} - 1) & = \frac{1}{2\pi}\int_{\mathbb{R}} dt_1\ \Exp\left[it_1\braket{\psi|\psi} \right] \Exp\left[- i t_1\right]
\\
\delta (\braket{\chi|\Lambda|\psi} - z') & = \delta \left(\frac{\braket{\chi|\Lambda|\psi} + \braket{\psi|\Lambda|\chi}}{2} - \Re z'\right) \delta \left(\frac{\braket{\chi|\Lambda|\psi} - \braket{\psi|\Lambda|\chi}}{2i} - \Im z'\right) \nonumber
\\ 
& = \frac{4}{\left( 2\pi \right)^2} \int_{\mathbb{R}} dt_2 dt_3\ \Exp \left[ i(t_2 - i t_3) \braket{\chi|\Lambda|\psi} + i(t_2 + i t_3) \braket{\psi|\Lambda|\chi}  \right] \Exp \left[ -2i \left(t_2 \Re z' + t_3 \Im z'\right)\right] \nonumber
\end{align}
\end{small}
with $z'=z \sqrt{\braket{\chi|\Lambda|\chi}}$. The second delta function comes with a complex valued argument and thus has to be understood as product of two delta functions concerning the real resp. the imaginary part. We also make use here of $\delta(a x) = \abs{a}^{-1}\delta(x)$, where the prefactor 4 stems from. Furthermore the transition probability in the numerator of \eqref{AIntegral} is written in terms of a generating function
\begin{equation} \label{AGeneratingFunction}
\abs{\braket{\chi |A \Lambda| \psi}}^{2n} = \frac{1}{i^n} \frac{\partial^n}{\partial J^n} \Exp \left[ iJ \braket{\chi |A \Lambda| \psi} \braket{\psi |\Lambda A^\dag| \chi} \right] \bigg|_{J=0}.
\end{equation}
The entire numerator of \eqref{AIntegral} is then
\begin{align} \label{ANumerator}
\frac{4}{\left(2\pi\right)^3 i}\frac{\partial}{\partial_J} \int d[\psi] d[t]\ &\Exp \left[ i\bra{\psi}\left(J \Lambda A^\dag\ket{\chi} \bra{\chi} A \Lambda + t_1 \unit \right)\ket{\psi} + i(t_2 - i t_3) \braket{\chi|\Lambda|\psi} + i(t_2 + i t_3) \braket{\psi|\Lambda|\chi}  \right] \nonumber
\\
&\Exp \left[ -2i(t_2 \Re z' + t_3 \Im z') - i t_1\right] \bigg|_{J=0}. 
\end{align}
with $d[t] = dt_1 dt_2 dt_3$ and the boundaries of integration like in \eqref{4FourierSubstitutions}. Now the Gaussian integration can be performed. We use the well known formula
\begin{equation}
\int_{\mathbb{C}^N} d^N\vc{x}\ \Exp \left[ i \vc{x} D \vc{x} + i \vc{S} \vc{x} + i \vc{x} \vc{S} \right] = \frac{(i \pi)^N}{\Det D} \Exp \left[ -i \vc{S}D^{-1}\vc{S}\right]
\end{equation}
for the Hermitian matrix $D$ and $\vc{S} \in \mathbb{C}^N$. In our case $D$ corresponds to $J \Lambda A^\dag\ket{\chi} \bra{\chi} A \Lambda + t_1 \unit$, which is Hermitian as it consists only of a projector and an identity, and $\vc{S}$ corresponds to $(t_2 + i t_3)\Lambda \ket{\chi}$. This yields
\begin{align}
\frac{4 (i \pi)^N}{\left(2\pi\right)^3 i}\frac{\partial}{\partial_J} \int d[t]\ &\frac{\Exp \left[-i \left( t_2^2 + t_3^2 \right) \braket{\chi|\Lambda \left( J \Lambda A^\dag \ket{\chi}\bra{\chi} A \Lambda + t_1 \right)^{-1} \Lambda|\chi} \right]}{\Det \left[ J \Lambda A^\dag\ket{\chi} \bra{\chi} A \Lambda + t_1 \right]}
\\
&\Exp \left[ -2i(t_2 \Re z + t_3 \Im z) - i t_1\right] \bigg|_{J=0}. \nonumber
\end{align}
The next step is taking the derivative. The derivative of a determinant dependent on a real parameter is
\begin{equation}
\frac{\partial}{\partial_J} \Det D_J = \Det D_J\ \Tr\left( D_J^{-1} \partial_J D_J\right)
\end{equation}
and therefore
\begin{align}
\frac{\partial}{\partial_J} & \left[ \frac{\Exp \left[ -i \left( t_2^2 + t_3^2 \right) \braket{\chi|\Lambda \left( J \Lambda A^\dag \ket{\chi}\bra{\chi}A \Lambda + t_1 \right)^{-1}\Lambda|\chi} \right]}{\Det \left[ J \Lambda A^\dag\ket{\chi} \bra{\chi} A \Lambda + t_1 \right]} \right]\bigg|_{J=0} = 
\\
\Exp & \left[ -i \frac{t_2^2 + t_3^2}{t_1} \braket{\chi|\Lambda^2|\chi} \right] \left[ -\frac{\braket{\chi|A \Lambda^2 A^\dag|\chi}}{t_1^{N+1}} + i \frac{t_2^2 + t_3^2}{t_1^{N+2}} \abs{\braket{\chi|A \Lambda^2|\chi}}^2 \right]. \nonumber
\end{align}
In the upcoming integrals we use the substitution $t_{2,3} \rightarrow \sqrt{\braket{\chi| \Lambda^2 |\chi}}\ t_{2,3}$ and $z' = z \sqrt{\braket{\chi|\Lambda^2|\chi}}$ like in the main text
\begin{small}
\begin{align}
\int d[t]\ \frac{1}{t_1^{N+1}}\ \Exp\left[-i \frac{t_2^2 + t_3^2}{t_1} \braket{\chi|\Lambda^2|\chi} -2 i \left( t_2 \Re z' + t_3 \Im z' \right) -it_1 \right] &= \frac{i^{N-1} \pi^2}{(N-1)!} \frac{\left( \abs{z}^2 - 1 \right)^{N-1}}{\braket{\chi|\Lambda^2|\chi}}
\\
\int d[t]\ \frac{t_2^2 + t_3^2}{t_1^{N+2}}\ \Exp\left[-i \frac{t_2^2 + t_3^2}{t_1} \braket{\chi|\Lambda^2|\chi} -2 i \left( t_2 \Re z' + t_3 \Im z' \right) -it_1 \right] &= \frac{i^{N} \pi^2}{(N-1)!} \frac{\left( \abs{z}^2 - 1 \right)^{N-1}}{\braket{\chi|\Lambda^2|\chi}^2}\nonumber
\\
& - \frac{i^{N - 1} \pi^2}{(N-2)!} \frac{\left( \abs{z}^2 - 1 \right)^{N-2}}{\braket{\chi|\Lambda^2|\chi}^2} \abs{z}^2. \nonumber
\end{align}
\end{small}
The integral in $t_2$ and $t_3$ is Gaussian again, whereas the $t_1$ integral can be reduced to
\begin{equation} \label{AFourierFraction}
\int_\mathbb{R}dt\ \frac{e^{i \alpha t}}{t^N} = \frac{i^N \pi}{(N-1)!} \alpha^{N-1}\ \sgn \alpha.
\end{equation}
We get to this result by writing $1/t^N$ as a derivative and using the differentiation rule of the Fourier transform
\begin{equation}
\mathcal{F} \left( \frac{1}{t^N} \right) = \frac{(-1)^{N-1}}{(N-1)!} \mathcal{F} \left( \partial_t^{N-1} \frac{1}{t} \right) = \frac{\left( i \alpha \right)^{N-1}}{(N-1)!} \mathcal{F} \left( \frac{1}{t} \right).
\end{equation}
Now we only have to set in
\begin{equation}
\mathcal{F} \left( \frac{1}{t} \right) = i \pi\ \sgn \alpha,
\end{equation}
which exists in the sense of its principal value \cite{Sneddon}. All in all the numerator \eqref{ANumerator} evaluates to
\begin{align} \label{ANumeratorResult}
& \int d[\psi] \abs{\braket{\chi |A \Lambda| \psi}}^2 \delta \left(\braket{\chi |\Lambda| \psi} - z' \right) \delta (\braket{\psi|\psi} -1) = 
\\
& \frac{\pi^N}{2\pi}\left[ \frac{\left( 1 - \abs{z}^2\right)^{N-1}}{(N-1)!} \left( \frac{\braket{\chi|AA^\dag|\chi}}{\braket{\chi|\Lambda^2|\chi}} - \frac{\abs{\braket{\chi|A|\chi}}^2}{\braket{\chi|\Lambda^2|\chi}^2} \right) + \frac{\left( 1-\abs{z}^2 \right)^{N-2} \abs{z}^2}{(N-2)!}\frac{\abs{\braket{\chi|A|\chi}}^2}{\braket{\chi|\Lambda^2|\chi}^2} \right]. \nonumber 
\end{align}
The denominator is calculated in the same manner. But as there is no generating function for the transition probability a derivative is not necessary
\begin{equation} \label{ADenominator}
\int d[\psi] \delta(\braket{\psi|\psi} - 1) \delta (\braket{\chi | \psi} - z') =  \frac{\pi^{N}}{2\pi \braket{\chi|\Lambda^2|\chi}} \frac{\left( 1 - \abs{z}^2 \right)^{N-2}}{\left( N - 2 \right)!}.
\end{equation}

We consider $\Lambda = \unit$ for now in order to make a geometrical interpretation. In this case it corresponds to the volume of the intersection of the unit sphere, described by $\braket{\psi | \psi} = 1$, and a cone with its symmetry axis along $\ket{\chi}$ and an aperture of $2 \theta = 2\ \Arccos\abs{z}$, described by $\braket{\chi|\psi} = z$. It vanishes for $\abs{z} = 1$ as the intersection will be merely a point. For $\abs{z} = 0$ we obtain the volume of the $2N -3$-dimensional unit sphere up to a prefactor of $1/4$
\begin{equation}
\frac{\pi^{N-1}}{2}\frac{1}{(N-2)!} = \frac{1}{4} \Vol S^{2N-3}.
\end{equation}
The origin of this prefactor lies in our choice for the arguments of the delta function, i.e. we chose $\delta (\braket{\psi|\psi} - 1)$ instead of $\delta (\norm{\psi} - 1)$. But this is taken care of in the normalization \eqref{ADenominator}. 

Returning to the general case and combining \eqref{ANumeratorResult} and \eqref{ADenominator} we finally arrive at
\begin{equation} \label{AHAFixedResult}
\HA_{\psi}\left[\abs{\braket{\chi |A\Lambda| \psi}}^2 \delta \left(\frac{\braket{\chi |\Lambda| \psi}}{\sqrt{\braket{\chi|\Lambda^2|\chi}}} - z\right)\right] = \frac{1 - \abs{z}^2}{N - 1} \braket{\chi|A \Lambda^2 A^\dag|\chi} + \frac{N\abs{z}^2 - 1}{N - 1} \frac{\abs{\braket{\chi|A \Lambda^2|\chi}}^2}{\braket{\chi|\Lambda^2|\chi}}.
\end{equation}
Likewise we can calculate the second moment of $\abs{\braket{\chi |A\Lambda| \psi}}^2$, which is necessary for the variance, by taking the second derivative in \eqref{AGeneratingFunction}. This yields
\begin{small}
\begin{align} \label{A2ndMoment}
\HA_{\psi}\left[\abs{\braket{\chi |A\Lambda| \psi}}^4 \delta \left(\frac{\braket{\chi |\Lambda| \psi}}{\sqrt{\braket{\chi|\Lambda^2|\chi}}} - z\right)\right] & = \frac{2 \left( 1 - \abs{z}^2 \right)^2}{N(N-1)} \braket{\chi|A \Lambda^2 A^\dag|\chi}^2
\\
& + \left[ \frac{2 \left( 1 - \abs{z}^2 \right)^2}{N(N-1)} - \frac{4 \left( 1 - \abs{z}^2 \right) \abs{z}^2}{N-1} + \abs{z}^4 \right] \frac{ \abs{\braket{\chi|A \Lambda^2|\chi}}^4 }{\braket{\chi|\Lambda^2|\chi}^2} \nonumber
\\
& - \left[ \frac{4 \left( 1 - \abs{z}^2 \right)^2}{N(N-1)} - \frac{4 \left( 1 - \abs{z}^2 \right) \abs{z}^2}{N-1} \right] \braket{\chi|A \Lambda^2 A^\dag|\chi} \frac{\abs{\braket{\chi|A \Lambda^2|\chi}}^2}{\braket{\chi|\Lambda^2|\chi}}. \nonumber
\end{align}
\end{small}

\subsection{Hilbert space average of positive integer powers of matrix elements} \label{Aaveragematrixelements}

In this appendix we will present a formula for the higher moments of matrix elements using the methods of \cite{Gemmer}. For $M$ (not necessarily) different matrices $B_m$ and uniformly distributed $\ket{\chi}$ we found
\begin{align} \label{AB1toBM}
\HA\left[ \braket{\chi|B_1|\chi} \cdots \braket{\chi|B_M|\chi} \right] & = \sum^N_{\forall m: i_m, j_m = 1} B_1^{i_1 j_1} \cdots B_M^{i_M j_M}\ \HA\left[\prod_{m=1}^M \chi^*_{i_m} \chi_{j_m} \right]
\\
& = \frac{(N-1)!}{(N+M-1)!} \sum^N_{\forall m: i_m, j_m = 1} B_1^{i_1 j_1} \cdots B_M^{i_M j_M}\ 
\Per{\begin{pmatrix}
\delta_{i_1 j_1} & \cdots & \delta_{i_M j_1}\\
\vdots &  & \vdots \\
\delta_{i_1 j_M} & \cdots & \delta_{i_M j_M}\\
\end{pmatrix}}, \nonumber
\end{align}
where the superscripts of the matrices denote their elements and $\Per$the permanent. In the first step we merely used the definitions of the Hilbert space average, making it possible to express $\ket{\chi}$ in an arbitrary basis with the coordinates $\chi_i = \braket{i|\chi}$. For convenience of the reader we give the explicit expressions for $M=2$ and $M=3$
\begin{align}
\HA\left[ \braket{\chi|B_1|\chi} \braket{\chi|B_2|\chi} \right] & = \frac{1}{N(N+1)} \left( \Tr\ B_1 B_2 + \Tr\ B_1 \Tr\ B_2 \right)
\\
\HA\left[ \braket{\chi|B_1|\chi} \braket{\chi|B_2|\chi} \braket{\chi|B_3|\chi} \right] & =  \frac{(N-1)!}{(N+2)!} ( \Tr\ B_1\ \Tr\ B_2\ \Tr\ B_3 + \Tr\ B_1 B_2\ \Tr\ B_3  \nonumber \\
& + \Tr\ B_1\ \Tr\ B_2 B_3 + \Tr\ B_1 B_3\ \Tr\ B_2 + \Tr\ B_1 B_2 B_3 + \Tr\ B_1 B_3 B_2 ). \nonumber
\end{align}
So \eqref{AB1toBM} contains the sum of all possible combinations of traces of the matrices $B_1,\cdots, B_M$. Thus the average of the product of $M$ different matrix elements consists of $M!$ terms. We carefully verified this formula up to $M=4$ by explicit calculation.

\subsection{Distribution of transition probabilities} \label{ADist}

We are interested in the distribution $p(s)$ of transition probabilities $\abs{\braket{\chi|U|\psi}}^2$ with uniformly distributed states and unitary $U$. We follow the same strategy as before, i.e. converting the upcoming integral into a Gaussian integral with subsequent Fourier transformation
\begin{align} \label{ADistDef}
p(s) & = \int_{\mathbb{C}^N} d[\chi] d[\psi] \delta\left( s - \abs{\braket{\chi|U|\psi}}^2 \right) \delta\left( \braket{\chi|\chi} -1 \right) \delta\left( \braket{\psi|\psi} -1 \right)
\\
& = \frac{1}{\left( 2\pi \right)^2} \int d[\chi] d[\psi] d[t] \Exp \left[ -i t_1 s  - i t_2 \right] \Exp \left[ i \bra{\chi}\left(t_1 U\ket{\psi} \bra{\psi}U^\dag + t_2 \right) \ket{\chi} \right] \delta\left( \braket{\psi|\psi} -1 \right) \nonumber
\\
& = \frac{\left(i \pi\right)^N}{\left( 2\pi \right)^2} \int d[\psi] d[t] \frac{\Exp \left[ -i t_1 s  - i t_2 \right]}{\Det \left[ t_1 U \ket{\psi} \bra{\psi} U^\dag + t_2  \right]} \delta\left( \braket{\psi|\psi} -1 \right). \nonumber
\end{align}
The determinant yields 
\begin{equation}
\Det\left[ t_1 U \ket{\psi} \bra{\psi} U^\dag + t_2  \right] = t_2^{N-1} \left( t_2 + t_1 \braket{\psi|\psi} \right).
\end{equation}
Thus we can apply the remaining $\delta$-function and arrive at the following Fourier integral
\begin{equation}
\frac{\left(i \pi\right)^N}{\left( 2\pi \right)^2} \int d[t] \frac{\Exp \left[ -i t_1 s  - i t_2 \right]}{t_2^{N-1} \left( t_2 + t_1 \right)} = \frac{\left(i \pi\right)^N}{\left( 2\pi \right)^2} \int d[t] \frac{\Exp \left[ -i t_1 s  + i \left( s -1\right) t_2 \right]}{t_2^{N-1} t_1},
\end{equation}
which can, after the substitution in the second step, be evaluated in terms of \eqref{AFourierFraction}. Thus, we find 
\begin{equation}
\frac{\pi^N}{4 \left( N-2 \right)!} \left( 1 - s\right)^{N-2},
\end{equation}
and are able to extract the $s$-dependence. We denote $\langle f(s) \rangle = \int_0^1 ds f(s) p(s)$. The normalization condition $\braket{1} = 1$ yields
\begin{equation} \label{APDF}
p(s) = \left( N-1 \right) \left( 1 - s \right)^{N-2},
\end{equation}
which is a probability distribution on $[0,1]$ of the Kumaraswamy type. The mean and the variance are
\begin{equation}
\langle s \rangle = \frac{1}{N} \qquad \text{and} \qquad \sigma^2 = \langle s^2 \rangle - \langle s \rangle^2 = \frac{N-1}{N^2 (N+1)} \sim \mathcal{O}\left( N^{-2} \right).
\end{equation}
Therefore it is $\langle s \rangle \sim \sigma$. Furthermore for the skewness and the kurtosis we find
\begin{equation}
\left\langle \left( \frac{s - \langle s \rangle}{\sigma} \right)^3 \right\rangle \xrightarrow[]{N \rightarrow \infty} 2  \qquad \text{and} \qquad \left\langle \left( \frac{s - \langle s\rangle}{\sigma} \right)^4 \right\rangle \xrightarrow[]{N \rightarrow \infty} 9.
\end{equation}
The probability density $\eqref{APDF}$ shares these properties with the exponential distribution $N e^{-Ns}$. In fact it converges to the exponential distribution for large $N$. Furthermore replacing the integration regime in \eqref{ADistDef} by $\mathbb{R}^N$ would yield the well known Porter-Thomas distribution.

\end{document}